\documentclass[preprints,article,accept,moreauthors,pdftex]{mdpi} 

\firstpage{1} 
\pubvolume{xx}
\issuenum{1}
\articlenumber{5}
\pubyear{2019}
\copyrightyear{2019}
\history{Received: date; Accepted: date; Published: date}

\Title{Evaluation of Granger causality measures for constructing networks from multivariate time series}

\Author{Elsa Siggiridou $^{1}$, Christos Koutlis $^{1,2}$, Alkiviadis Tsimpiris $^{3}$ and Dimitris Kugiumtzis $^{1,}$*}

\AuthorNames{Elsa Siggiridou, Christos Koutlis, Alkiviadis Tsimpiris and Dimitris Kugiumtzis}

\address{%
$^{1}$ \quad Department of Electrical and Computer Engineering, Aristotle University of Thessaloniki, University Campus, 54124, Thessaloniki, Greece; E-mail: dkugiu@auth.gr; esingiri@auth.gr\\
$^{2}$ \quad Information Technologies Institute, Centre of Research and Technology Hellas, 57001, Thessaloniki, Greece;  E-mail: ckoutlis@iti.gr\\
$^{3}$ \quad Department of Computer, Informatics and Telecommunications Engineering,
International Hellenic University, 62124, Serres, Greece; E-mail: atsimpiris@teicm.gr\\}

\corres{Author to whom correspondence should be addressed; E-Mail: dkugiu@auth.gr;
Tel: +30-2310995955.}

\abstract{Granger causality and variants of this concept allow the study of
complex dynamical systems as networks constructed from
multivariate time series. In this work, a large number of Granger causality
measures used to form causality networks from multivariate time
series are assessed. These measures are in the time domain, such
as model-based and information measures, the frequency domain and
the phase domain. The study aims also to compare bivariate and
multivariate measures, linear and nonlinear measures, as well as
the use of dimension reduction in linear model-based measures and
information measures. The latter is particular relevant in the
study of high-dimensional time series. For the performance of the
multivariate causality measures, low and high dimensional coupled
dynamical systems are considered in discrete and continuous time,
as well as deterministic and stochastic. The measures are
evaluated and ranked according to their ability to provide
causality networks that match the original coupling structure. The
simulation study concludes that the Granger causality measures using
dimension reduction are superior and should be preferred
particularly in studies involving many observed variables, such as
multi-channel electroencephalograms and financial markets.}


\begin{document}


\section{Introduction}
Real-world complex systems have been studied as networks formed
from multivariate time series, i.e. observations of a number of
system variables, such as financial markets and brain dynamics
\cite{Arenas08,Zanin12b}. The nodes in the network are the
observed variables and the connections are defined by an
interdependence measure. The correct estimation of the
interdependence between the observed system variables is critical for the
formation of the network and consequently the identification of the
underlying coupling structure of the observed system.

Many interdependence measures that quantify the causal effect
between the variables observed simultaneously in a time series are
based on the concept of Granger causality \cite{Granger69}: a
variable $X$ Granger causes a variable $Y$ if the information in
the past of $X$ improves the prediction of $Y$. 
The concept was first mentioned by Wiener \cite{Wiener56} 
and it is also referred to as Wiener-Granger causality, but for brevity we 
use the common term Granger causality here.
The methodology on
Granger causality was first developed in econometrics and it has
been widely applied to many other fields, such as cardiology and
neuroscience (analysis of electroencephalograms (EEG) and
functional magnetic resonance imaging (fMRI)
\cite{Smith11,Bastos16,Porta16,Fornito16}) and climate
\cite{Hlinka11,Dijkstra19}.

The initial form of Granger causality based on autoregressive
models has been extended to nonlinear models, basically local
linear models \cite{Schiff96,LeVanQuyen99b,Chen04,Faes08}, but
also kernel-based and radial basis models
\cite{Marinazzo06,Marinazzo08,Marinazzo11,Karanikolas17}, and
recently more advanced models, such as neural networks
\cite{Montalto15,Abbasvandi19}. In a wider sense, the directed
dependence inherent in Granger causality is referred to as
coupling, synchronization, connectivity and information flow
depending on the estimation approach for the interdependence.
Geweke \cite{Geweke82,Geweke84} defined an analogue of Granger
causality in the frequency domain, developed later to other
frequency measures \cite{Kaminski91,Baccala01}. A number of
nonlinear measures of interdependence inspired by the Granger
causality idea have been developed, making use of state-space
techniques \cite{Arnhold99,Romano07}, information measures
\cite{Schreiber00,Palus96,Vlachos10}, and techniques based on the
concept of synchronization \cite{Rosenblum01,Nolte08,Lehnertz09}.
We refer to all these measures as causality measures (in the
setting of multivariate time series) and the networks derived by
these measures as causality networks.

Distinguishing indirect and direct causality with the available
methods is a difficult task \cite{Rings2016,Han2016}, and
multivariate measures are expected to address better this task
than bivariate measures. The bivariate causality measures do not
make use of the information of other observed variables besides
the variables $X$ and $Y$ of the examined causality from $X$ to
$Y$, denoted $X \rightarrow Y$, and thus estimate direct and
indirect causality, where indirect causality is this mediated by a
third variable $Z$, e.g. $X \rightarrow Z$ and $Z \rightarrow Y$
results in $X \rightarrow Y$. The multivariate causality measures
apply conditioning on the other observed variables to estimate
direct causal effects, denoted as $X \rightarrow Y | Z$, where $Z$
stands for the other observed variables included as conditioning
terms \cite{Blinowska04,Kus04,Eichler12}.

For high-dimensional time series, i.e. large number $K$ of
observed variables, the estimation of direct causal effects is
difficult and the use of multivariate causality measures is
problematic. One solution to this problem is to account only for a
subset of the other observed variables based on some criterion of
relevance to the driving or response variable \cite{Marinazzo12}.
In a different approach, dimension reduction techniques have been
embodied in the computation of the measure restricting the
conditioning terms and they have been shown to improve the
efficiency of the direct causality measures
\cite{Vlachos10,Faes11,Runge12,Wibral12,Kugiumtzis13,Siggiridou16}.

Recent comparative studies have assessed causal effects with various causality measures, using also
significance tests for each causal effect
\cite{Smirnov05,Faes08,Papana11,Silfverhuth12,Fasoula13,Papana13,Zaremba14,Cui16,Bakhshayesh19}.
Some studies concentrated on the comparison of direct and indirect
causality measures \cite{Blinowska11,Silfverhuth12}, whereas other
studies focused on specific types of causality measures, e.g.
frequency domain measures
\cite{Astolfi05,Florin11,Wu11,Sommariva19}, or different
significance tests for a causality measure
\cite{Diks17,Papana17,Moharramipour18}. These studies are done on
specific real data types, mostly from brain, which limits the
generalization of the conclusions.

Whereas most of the abovementioned studies were concentrated on
the estimation of specific causal effects by the tested measures,
this study is merely focused on assessing the bivariate ($X
\rightarrow Y$) and multivariate and ($X \rightarrow Y | Z$) causality 
measures that estimate best the whole set of
causal effects for all pairs ($X$,$Y$), i.e. the true causality
network. In particular, high dimensional systems and subsequently
high-dimensional time series are considered, so that the estimated
networks have up to 25 nodes. Some first results of the
application of different causality measures on simulated systems
and evaluation of their accuracy in matching the original network
were presented earlier in \cite{Siggiridou15}. As the focus is on
the preservation of the original causality network, we assess the
existence of each causality term applying appropriate significance 
criteria. The causality measures are ranked as to their
ability to match the original causality networks of different
dynamical systems and stochastic processes. For the computation of
the causality measures, several software are freely accessible
\cite{Cui08,Seth10,Niso13,Lizier14,Barnett14,Montalto14}, but we
have developed most of the causality measures in the context of
previous studies of our team, and few measures were run from the
software \cite{Niso13}.

The structure of the paper is as follows. In
Sec.~\ref{sec:Methods}, we present the causality measures, the
identification of network connections from each measure, the
statistical evaluation of the accuracy of each measure in
identifying the original coupling network, the formation of a
score index for each measure, and finally the synthetic systems
used in the simulations. In Sec.~\ref{sec:Results}, we present the
results of the measures on these systems, and we rank the measures
as for their accuracy in matching the original coupling network.
Discussion and conclusions follow in Sec.~\ref{sec:Discussion}.

\section{Methodology}
\label{sec:Methods}

The methodology implemented for the comparative study of causality
measures aiming at evaluating the measures and ranking them as to
their accuracy in identifying correctly the underlying coupling
network is presented here. The methodology includes the causality
measures compared in the study, the techniques for the
identification of network connections from each measure, the
statistical evaluation of the accuracy of each measure in
identifying the original coupling network, and the formation of an
appropriate score index for the overall performance of each
measure. Finally, the synthetic systems used in the simulation
study are presented.

\subsection{Causality measures}
\label{subsec:measures}

First, it is noted that in this comparative study correlation or
in general symmetric measures of $X$ and $Y$ are not considered.
Many such measures were initially included in the study, e.g. many
phase-based measures such as phase locking value
(PLV)\cite{Lachaux99}, phase lag index  (PLI) \cite{Stam07} and
weighted phase lag index (wPLI) \cite{Vinck11}, rho index (RHO)
\cite{Tass98}, phase slope index (PSI) \cite{Nolte08} and mean
phase coherence (MPC) \cite{Mormann00}. However, their evaluation
in the designed framework is not possible as the identification of
the exact directed connections of the original coupling network is
quantified to assess the measure performance.

\begin{figure}[h!]
    \centering
    \includegraphics[scale=0.7]{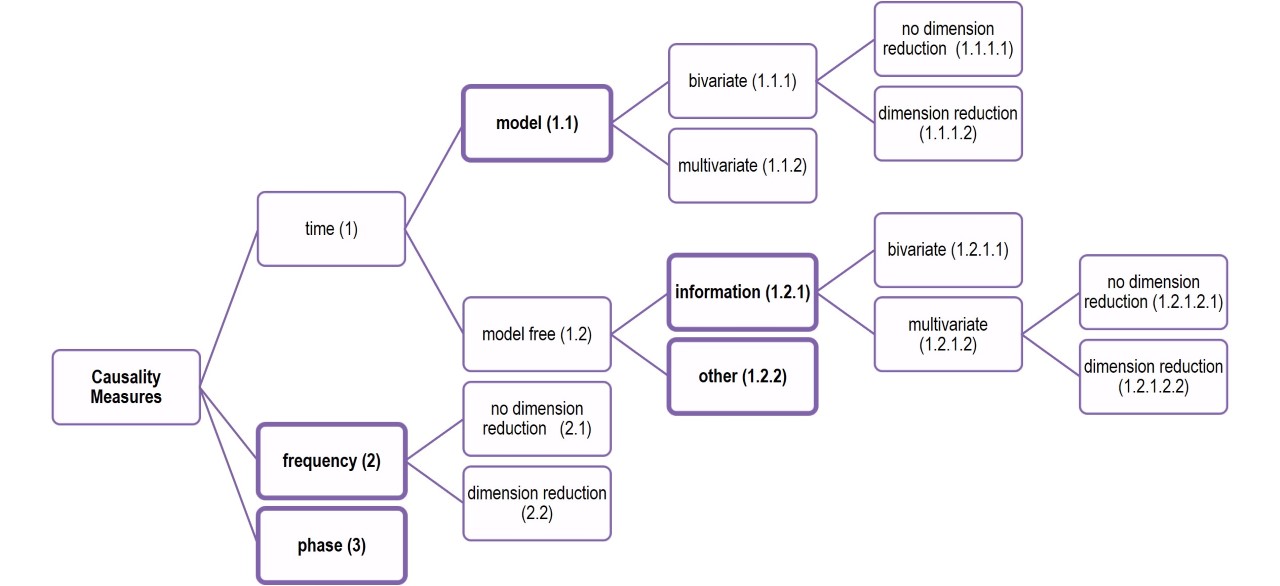}
    \caption{Tree structure for the types of causality measures.
    The five main classes are highlighted (frame box in bold).
    Each measure type in a box is given a code number, used as
    reference in Table~\ref{tab:MeasureList}.}
    \label{fig:SkecthMeasures}
\end{figure}

Causality measures can be divided in three categories as to the
domain of data representation they are defined in: time, frequency
and phase (see Fig.\ref{fig:SkecthMeasures}).
The measures in time domain dominate and they are
further divided in model-based and model-free measures. Many of
the model-free measures are based on information theory measures
and the other model-free measures on the time domain are referred
to as ``other'' measures. Thus five main classes of causality
measures are considered in this study: model-based measures,
information measures, frequency measures, phase measures and other
measures that cannot be defined in terms of the other four
classes. The measures organized in these five classes and included
in the comparative study are briefly discussed below, and they are
listed in Table~\ref{tab:MeasureList}, with reference and code
number denoting the type of measure (the class, bivariate or
multivariate and with our without dimension reduction).

\begin{table}[H]
\caption{List of causality measures organized in the five classes.
The first column has the measure notation including the measure
parameters, the second column has a brief description, the third
column has the code of each measure denoting its type, and the
fourth column has a related reference.}
\centering
\begin{tabular}{c p{9cm} cc}
\toprule
\textbf{Symbol} & \textbf{Description}  & \textbf{Type} & \textbf{Ref.}\\
\midrule
\multicolumn{4}{c}{\textit{Model-based}}\\
\midrule
GCI($p$)        & Granger causality index, $p$ is the VAR order, for H\'{e}non maps $p\!=\!2,5$, for VAR
process $p\!=\!3,5$, for Mackey-Glass system and neural mass model
$p\!=\!5,10,20$
 & 1.1.1.1 & \cite{Granger69} \\
CGCI($p$)          & Conditional Granger causality index & 1.1.2 & \cite{Geweke84} \\
PGCI($p$)          & Partial Granger causality index & 1.1.2 & \cite{Guo08} \\
RCGCI($p$)        &  Restricted Granger causality index  & 1.1.1.2 & \cite{Siggiridou16} \\
\midrule
\multicolumn{4}{c}{\textit{Information}} \\
\midrule
TE($m,\tau$)    & Transfer entropy. Lag $\tau\!=\!1$ for all
systems, embedding dimension for H\'{e}non maps $m\!=\!2,3$, for VAR
process $m\!=\!3,5$, for Mackey-Glass system and neural mass model
$m\!=\!5,10,15$ & 1.2.1.1& \cite{Schreiber00} \\
PTE($m,\tau$)   & Partial transfer entropy &1.2.1.2.1 & \cite{Papana12} \\
STE($m,\tau$)   & Symbolic transfer entropy & 1.2.1.1&
\cite{Staniek08} \\
PSTE($m,\tau$)  & Partial symbolic transfer entropy &1.2.1.2.1 & \cite{Papana15} \\
TERV($m,\tau$)  & Transfer entropy on rank vectors &1.2.1.1 & \cite{Kugiumtzis12} \\
PTERV($m,\tau$) & Partial transfer entropy on rank vectors &1.2.1.2.1 & \cite{Kugiumtzis13b} \\
PMIME($L$)      & Partial mutual information from mixed embedding,
maximum lag for H\'{e}non maps and VAR model $L_{max}\!=\!5$, for
Mackey-Glass system and neural mass model $L_{max}\!=\!20$ &
1.2.1.2.2& \cite{Kugiumtzis13}
\\
\midrule
\multicolumn{4}{c}{\textit{Frequency}}\\
\midrule
PDC($p,i$)      & Partial directed coherence, $i$ denotes
the power band $i\!=\!\delta,\theta,\alpha,\beta,\gamma$ (relative
proportion of the whole spectrum). For H\'{e}non maps $p\!=\!2,5$, for the
VAR process $p\!=\!3,5$, for the Mackey-Glass system and the neural mass model
$p\!=\!5,10,20$ & 2.1 & \cite{Baccala01} \\
GPDC($p,i$)       & Generalized partial directed coherence & 2.1 & \cite{Baccala07}\\
DTF($p,i$)        & Directed transfer function & 2.1 & \cite{Kaminski91} \\
dDTF($p,i$)       & Direct directed transfer function & 2.1 & \cite{Korzeniewska03} \\
GGC($p,i$)        & Geweke's spectral Granger causality & 1.1.1.1 & \cite{Hu11} \\
RGPDC($p,i$)       & Restricted generalized partial directed coherence  & 2.2 & \cite{Siggiridou17} \\
\midrule
\multicolumn{4}{c}{\textit{Phase}} \\
\midrule
DPI             & Phase directionality index & 3 & \cite{Rosenblum01} \\
\midrule
\multicolumn{4}{c}{\textit{Other}} \\
\midrule
MCR($m,\tau$)   & Mean conditional recurrence, $m$ is the same as for the information measures & 1.2.2& \cite{Romano07} \\
DED             & Directed event delay & 1.2.2& \cite{QuianQuiroga02} \\
\label{tab:MeasureList}\\
\bottomrule
\end{tabular}
\end{table}

The first class of model-based measures regards measures that
implement the original concept of Granger causality, the bivariate
measure of Granger causality index (GCI) \cite{Granger69} (only
$X$ and $Y$ variables are considered), and the multivariate
measures of the conditional Granger causality index (CGCI)
\cite{Geweke84} and the partial Granger causality (PGCI)
\cite{Guo08} (also the other observed variables denoted $Z$ are
included). All these measures require the fit of a vector
autoregressive model (VAR), on the two or more variables. 
The order $p$ of VAR denoting the lagged variables of each variable contained 
in the model can be estimated by an information criterion such as AIC or BIC, which
often does not provide optimal lags, e.g. see the simulation study in \cite{Hatemi-J09} and citations therein, and the so-called $p$-hacking (in the sense of $p$-value) in terms of VAR models for Granger causality in \cite{Bruns19}. To overcome the use of order estimation criteria, here we use a couple of predefined appropriate orders $p$ for each system (see Table~\ref{tab:MeasureList}).
In the presence of many observed variables, dimension reduction in VAR
has been proposed, and here we use the one developed from our
team, the restricted conditional Granger causality index (RCGCI)
\cite{Siggiridou16}. Thus for this class of measures, there are
bivariate and multivariate measures, and multivariate measures
that apply dimension reduction, as noted in the sketched division
of causality measures in Fig.\ref{fig:SkecthMeasures}. These are
all linear measures and besides this constraint they have been
widely used in applications. Other nonlinear extensions are not
considered in this study for two reasons: either they were very
computationally intensive, such as the cross predictions of local
state space models \cite{Faes08}, or they were too complicated so
that discrepancies to the original methods may occur
\cite{Montalto15,Abbasvandi19}.

The information measures of the second class have also been
popular in diverse applications recently due to their general
form, as they do not require any specific model, they are
inherently nonlinear measures and can be applied to both
deterministic systems and stochastic processes of any type, e.g.
oscillating flows and discrete maps of any dimension. The main
measure they rely on is the mutual information (MI), and more
precisely the conditional mutual information (CMI). There have
been several forms for causality measures based on MI and CMI in
the literature, e.g. see the coarse-grained mutual information in
\cite{Palus01b}, but the prevailing one is the transfer entropy
(TE), originally defined for two variables \cite{Schreiber00}. The
multivariate version, termed partial transfer entropy (PTE) was
later proposed together with different estimates of the entropies
involved in the definition of PTE, bins \cite{Verdes05},
correlation sums \cite{Vakorin09} and nearest neighbors
\cite{Papana12}. Here, we consider the nearest neighbor estimate
for both TE and PTE, found to be the most appropriate for high
dimensions. Equivalent forms to TE and PTE are defined for the
ranks of the embedding vectors rather than the observations
directly. We consider the bivariate measures of symbolic
transfer entropy (STE) \cite{Staniek08} and transfer entropy on
rank vectors (TERV) \cite{Kugiumtzis12}, and the multivariate
measures of partial symbolic transfer entropy (PSTE)
\cite{Papana15} and partial transfer entropy on rank vectors
(PTERV) \cite{Kugiumtzis13b}. The idea of dimension reduction was
implemented in TE first, applying a scheme for a sparse
non-uniform embedding of both $X$ and $Y$, termed mutual
information on mixed embedding (MIME) \cite{Vlachos10}. This
bivariate measure was later extended to the multivariate measure
of partial MIME (PMIME) \cite{Kugiumtzis13}. Only the PMIME is
included in the study simply due to the computational cost, and it
is noted that by construction the measure gives zero for
insignificant causal effects, so it does not require binarization
when networks of binary connections have to be derived (the
positive values are simply set to one).

All the methods in the third class of frequency measures rely on
the estimation of the VAR model, either on only the two variables
$X$ and $Y$ or on all the observed variables (we consider only the
latter case here). Geweke's spectral Granger causality (GGC) is
the early measure implementing the concept of Grangre causality in
the frequency domain \cite{Geweke82,Hu11}, included in the study.
Another older such measure included in the study is the direct
transfer function (DTF) \cite{Kaminski91}, which although it is 
multivariate measure it does not discriminate direct from indirect
causal effects. For this, an improvement is proposed and used also
in this study, termed direct directed transfer function (dDTF)
\cite{Korzeniewska03}. We also consider the partial directed
coherence (PDC) \cite{Baccala01} and the improved version of
generalized partial directed coherence (GPDC) \cite{Baccala07},
which have been particularly popular in EEG analysis. When applied 
to EEG, the measures are defined in terms of frequency bands of
physiological relevance ($\delta$, $\theta$, $\alpha$, $\beta$, $\gamma$,
going from low to high frequencies), and the same proportional split of the
frequency range is followed here (as if the sampling frequency was 100 Hz). 
Finally, we consider a dimension reduction of VAR in the GPDC measure called
restricted GPDC (RGPDC), recently proposed from our team
\cite{Siggiridou17}.

As mentioned above the class of phase measures contains a good
number of measures used in connectivity analysis, mainly in
neuroscience dealing with oscillating signals such as EEG, but
most of these measures are symmetric and thus out of the scope of
the current study. In the evaluation of the causality measures we
consider the bivariate measure of phase directionality index (DPI)
\cite{Rosenblum01}, which is a measure of synchronization designed
for oscillating time series. Information measures have also been
implemented in the phases, e.g. see \cite{Palus03}, but not
considered here.

Another class of measures used mainly in neuroscience regards
inter-dependence measures based on neighborhoods in the
reconstructed state space of each of the two variables $X$ and
$Y$. A series of such measures have been proposed after the first
work in \cite{Arnhold99}, using also ranks of the reconstructed
vectors \cite{Chicharro09}, the latter making the measure
computationally very slow. The convergent cross mapping is
developed under the same approach \cite{Sugihara12}, and the same
yields for the measure of mean conditional recurrence (MCR)
\cite{Romano07}. It is noted that all these measures are bivariate
and they are expected to suffer from estimating indirect
connections in the estimated causality network. The MCR is
included as a representative of the state space bivariate measures
in the class of other measures. In this class, we include also
event synchronization measures \cite{QuianQuiroga02}, and
specifically the direct event delay (DED) that is a directional
bivariate measure, considered as causality measure and included in
the study.

For the information measures where the estimation of entropies in high
dimensions is hard, the comparison of the multivariate measures that do not include
dimension reduction to these including dimension reduction would
be unfair when high-dimensional systems are considered. To address
this, in the calculation of a multivariate information measure not making use
of dimension reduction, we choose to restrict the set of the
conditioning variables $Z$ in the estimated causal effect $X
\rightarrow Y|Z$ to only the three more relevant variables. In the
simulations, we consider the number of observed variables (equal to
the subsystems being coupled) to be $K\!=\!5$ and $K\!=\!25$. While for
$K\!=\!5$ all the remaining variables are considered in $Z$, for
$K\!=\!25$ only three of the remaining 23 variables are selected. The
criterion of selection is the mutual information of the remaining
variables to the driving variable $X$, which is common for the
selection of variables \cite{Marinazzo12}.

\subsection{Identification of original network connections}
\label{subsec:Network}

We suppose that a dynamical system is formed by the coupling of
$K$ subsystems and we observe one variable from each subsystem, so
that a multivariate time series of dimension $K$ is derived. The
coupling structure of the original system can be displayed as a
network of $K$ nodes where the connections are determined by the
system equations. Formally, in the graph-theoretical framework, a
network is represented by a graph $G \!=\! (V;E)$, where $V$ is the
set of $K$ nodes, and $E$ is the set of the connections among the
nodes of $V$. The original coupling network is given as a graph of directed
binary connections, where the connection from node $i$ to node $j$
is assigned with a value $a_{i,j}$ being one or zero, depending
whether variables of the subsystem $i$ are present in the equation
determining the variables of the subsystem $j$. The components
$a_{i,j}$, $i,j\!=\!1,\ldots,K$, form the adjacency matrix $A$ that
defines the network. 

The computation of any causality
measure presented in Sec.~\ref{subsec:measures} on all the
directed pairs $(i,j)$ of the $K$ observed variables gives a
weight matrix $R$ (assuming only positive values of the measure,
so that a transformation of the measure can be applied if
necessary). Thus, the pairwise causality matrix $R$ with entries
$R_{i,j} \!=\! R_{X_{i} \rightarrow X_{j}}$ defines a network of
weighted connections, assigning the weighted directed connection
$R_{i,j}$ from each node $i$ to each node $j$.

In applications, often binary networks are sought to better
represent the estimated structure of the underlying system. Here,
we are interested to compare how the causality measure retrieves
the original directed coupling structure, and therefore we want to
transform the weighted network to a binary network. Commonly, the
weighted matrix $R$ is transformed into an adjacency matrix $A$ by
suitable thresholding, keeping in the graph only connections with
weights higher than some threshold (and setting their weights to
one) and removing the weaker connections (setting their weights to
zero). For each causality measure, an appropriate threshold
criterion is sought to determine the significant values of the
measure that correspond to connections in the binary
network. We consider three approaches for thresholding that have
been used in the literature.
\begin{enumerate}[leftmargin=*,labelsep=4.9mm]
\item Rigorous thresholding is provided by appropriate
significance test for the causality measure $R_{X_{i} \rightarrow
X_{j}}$. For all considered causality measures in this study, we
expect the causality measure to lie at the zero level if there is
no causal effect and be positive if it is, so that the test is
one-sided. So, the null and alternative hypotheses are
respectively:
\begin{equation}
\mbox{H}_{0} : R_{X_{i} \rightarrow X_{j}}=0, \quad\quad
\mbox{H}_{1} : R_{X_{i} \rightarrow X_{j}} >0
\label{eq:Hypotheses}
\end{equation}

Parametric significance tests have been developed only for the
linear causality measures, and for consistency we apply the randomization
significance test to all causality measures, making use of the
simple technique of time-shifted surrogates. Specifically, we
generate $M$ resampled (surrogates) time series for the driving
variable $X$, each by shifting cyclically the original
observations of $X$ by a random step $w$. For the original time series of the
driving variable $X_i$ denoted $\{X_{i,t}\} \!=\! \{x_{1,t},x_{2,t},\ldots,x_{n,t}\}$, the
surrogate time series is $\{X_{i,t}^*\} \!=\! \{x_{w+1,t}, x_{w+2,t},
\ldots, x_{n,t}, x_{1,t}, \ldots, x_{w-1,t}, x_{w,t}\}$. In this
way we destroy any form of coupling of $X_i$ and any other
variable $X_j$, so that $\{X_{i,t}^*\}$ is consistent to $H_{0}$,
but it preserves the dynamics and the marginal distribution of
$X_i$. The test statistic is the causality measure $R_{X_{i}
\rightarrow X_{j}}$, and it takes the value $R_{0}$ on the
original time series pair and the values
$R_{1},R_{2},\ldots,R_{M}$ on each of the $M$ resampled time
series pairs. The rank of $R_{0}$ in the ordered list of $M+1$
values $R_{0},R_{1},R_{2},\ldots,R_{M}$, denoted $r_{0}$, defines
the $p$-value of the randomization test as
$p\!=\!1-\frac{r_{0}-0.326}{M+1+0.348}$ \cite{Yu01}. If $R_{0}$ is at
the right tail of the empirical distribution formed by
$R_{1},R_{2},\ldots,R_{M}$, then the H$_{0}$ is rejected, which
suggests that $p<\alpha$, where $\alpha$ is the significance level
of the test determining the tail. For a multivariate time series
of $K$ variables, totally $K(K-1)$ significance tests are
performed for each causality measure, indicating that multiple
tests are performed on the same dataset. This is a known issue in
statistics and corrections for multiple testing can be further be
applied, such as the false discovery rate \cite{Benjamini95}.
Here, we refrain from using such a correction and rather use three
different significance levels $\alpha\!=\!0.01, 0.05, 0.1$. We opted
for this as the same setting is applied for all causality
measures.

\item The second thresholding criterion is given by the desired
density of the binary network. In the simulation study, we know
the density of the original network, denoted by the number of
connections $\rho_0$. We consider five different values for the
density $\rho$ of the estimated causality binary network given in
multiples of $\rho_0$ as $0.6,0.8,1.0,1.2,1.4$.

\item The third thresholding criterion is simply given by a
predefined magnitude threshold on the causality measure. Here, we
select an appropriate threshold $th_{\rho}$ separately for each
causality measure and each coupling strength for the same system,
where $\rho$ indicates the corresponding density. Having 10
realizations for each scenario (system and coupling strength), the
$th_{\rho}$ is the average of the thresholds found for the given
density $\rho$ in the 10 realizations.
\end{enumerate}

\subsection{Performance indices for statistical evaluation of methods accuracy}
\label{subsec:accuracy}

For a system of $K$ variables there are $K(K-1)$ ordered pairs of
variables to estimate causality. In the simulations of known
systems, we know the true coupling pairs and thus we can compute
performance indices to rate the causality measures as for their
overall matching of the original connections in the network. Here,
we consider the performance indices of specificity, sensitivity, Matthews
correlation coefficient, F-measure and Hamming distance.

The sensitivity is the proportion of the true causal effects (true
positives, TP) correctly identified as such, given as
sens=TP/(TP+FN), where FN (false negatives) denotes the number of
pairs having true causal effects but have gone undetected. The
specificity is the proportion of the pairs correctly not
identified as having causal effects (true negatives, TN), given as
spec=TN/(TN+FP), where FP (false positives) denotes the number of
pairs found falsely to have causal effects. An ideal causality
measure would give values of sensitivity and specificity at one.
To weigh sensitivity and specificity collectively we consider the
Matthews correlation coefficient (MCC) \cite{Matthews75} given as
\begin{equation}
 \text{MCC}=\frac{\text{TP}\cdot \text{TN}-\text{FP}\cdot \text{FN}}
{\sqrt{(\text{TP}+\text{FP})\cdot (\text{TP}+\text{FN})\cdot 
(\text{TN}+\text{FP})\cdot (\text{TN}+\text{FN})}}.
 \label{eq:MCCdef}
\end{equation}
MCC ranges from -1 to 1. If MCC=1 there is perfect identification
of the pairs of true and no causality, if MCC=-1 there is total
disagreement and pairs of no causality are identified as pairs of
causality and vice versa, whereas MCC at the zero level indicates
random assignment of pairs to causal and non-causal effects.

Similarly, we consider the F-measure that combines precision and
sensitivity. The precision, called also positive predictive value,
is the number of detected true causal effects divided by the total
number of detected casual effects, given as prec=TP/(TP+FP), and
the F-measure (FM) is defined as
\[
\mbox{FM} = \frac{2\cdot \text{prec} \cdot \text{sens}}{\text{prec} +
\text{sens}} = \frac{2\text{TP}}{2\text{TP}+\text{FN}+\text{FP}},
\]
which ranges from 0 to 1. If FM=1 there is perfect identification
of the pairs of true causality, whereas if FM=0 no true coupling
is detected.

The Hamming distance (HD) is the sum of false positives (FP) and false
negatives (FN), HD=FP+FN. Thus HD gets non-negative integer values
bounded below by zero (perfect identification) and above by
$K(K-1)$ if all pairs are misclassified.

\subsection{Score Index}
\label{subsec:score}

In Sec.~\ref{subsec:accuracy}, we presented five performance
indices to evaluate in different ways the match of the original
network and the estimated network from each causality measure.
Further, we want to quantify this match for different settings,
which involve different systems and different scenarios for each
system regarding the number of variables $K$, the system free
parameter, the time series length $n$ and the coupling strength $C$,
where applicable (to be presented below in Sec.~\ref{subsec:Systems}). 
For this, we rely on the index MCC, and for each
scenario of a different system, we rank the causality measures
according to their average MCC (across 10 realizations generated per scenario). For equal MCC, ordinal ranking (called
also $"1234"$ ranking) is adopted \cite{Lansdown98}. Specifically,
the order of measures of equal MCC is decided from distinct
ordinal numbers given at random to each measure of equal MCC
value. Next, we derive the average rank of a causality measure $i$
for all different coupling strengths $C$ of a system $j$,
$P_{i,j}$, as the average of the ranks of the causality measure $i$ in all
coupling strengths tested for the system $j$. A score $s_{i,j}$ of
the causality measure $i$ for the system $j$ is then derived by
normalization of the average rank $P_{i,j}$ over the number $N$ of
all measures so as to scale between zero and one,
$s_{i,j}\!=\!(N-P_{i,j})/(N-1)$, where one is for the best measure ranked at top. 
The overall score of the causality
measure $i$ over all systems, $s_i$, is simply given by the
average $s_{i,j}$ over all systems, including the two different
$K$ values for systems S1, S2 and S3, the two values for the
system parameter $\Delta$ and $A$ for systems S2 and S3,
respectively, and the two time series lengths $n$ for system S1.
The systems are presented in the next section.

\subsection{Synthetic systems}
\label{subsec:Systems}

For the comparative study, we use four systems with different
properties: the coupled H\'{e}non maps as an example of
discrete-time chaotic coupled system \cite{Kugiumtzis13a}, the
coupled Mackey-Glass system as an example of continuous-time
chaotic coupled system but of high complexity
\cite{Senthilkumar08,Kugiumtzis13a}, the so-called neural mass
model as an example of a continuous-time stochastic system used as
an EEG model \cite{Wendling00,Koutlis16}, and the vector
autoregressive model (VAR) as suggested in \cite{Basu15}, used as
an example of a discrete-time linear stochastic process. The four
systems are briefly presented below.

\begin{figure}[H]
\centering
\includegraphics[scale=0.25]{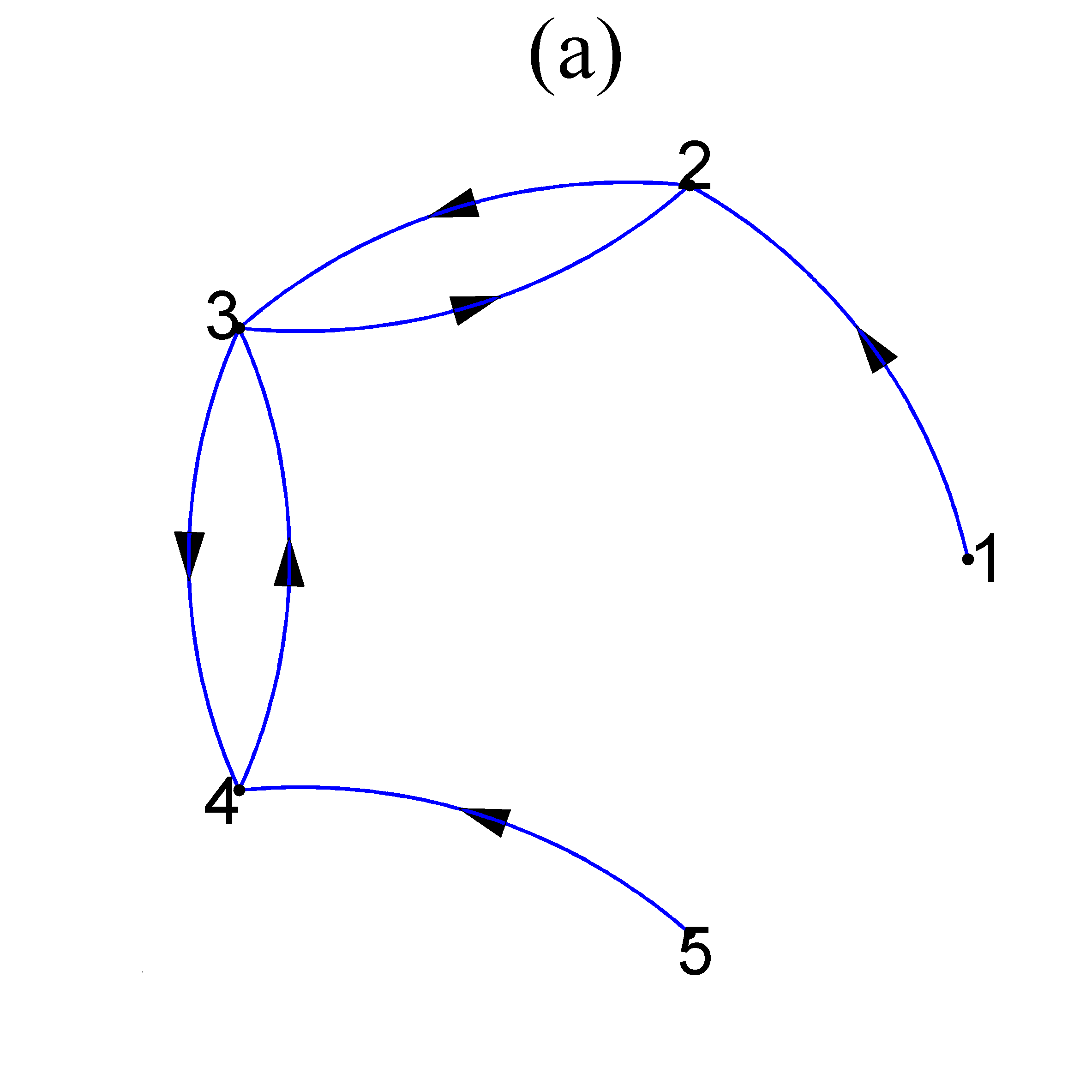}
\includegraphics[scale=0.25]{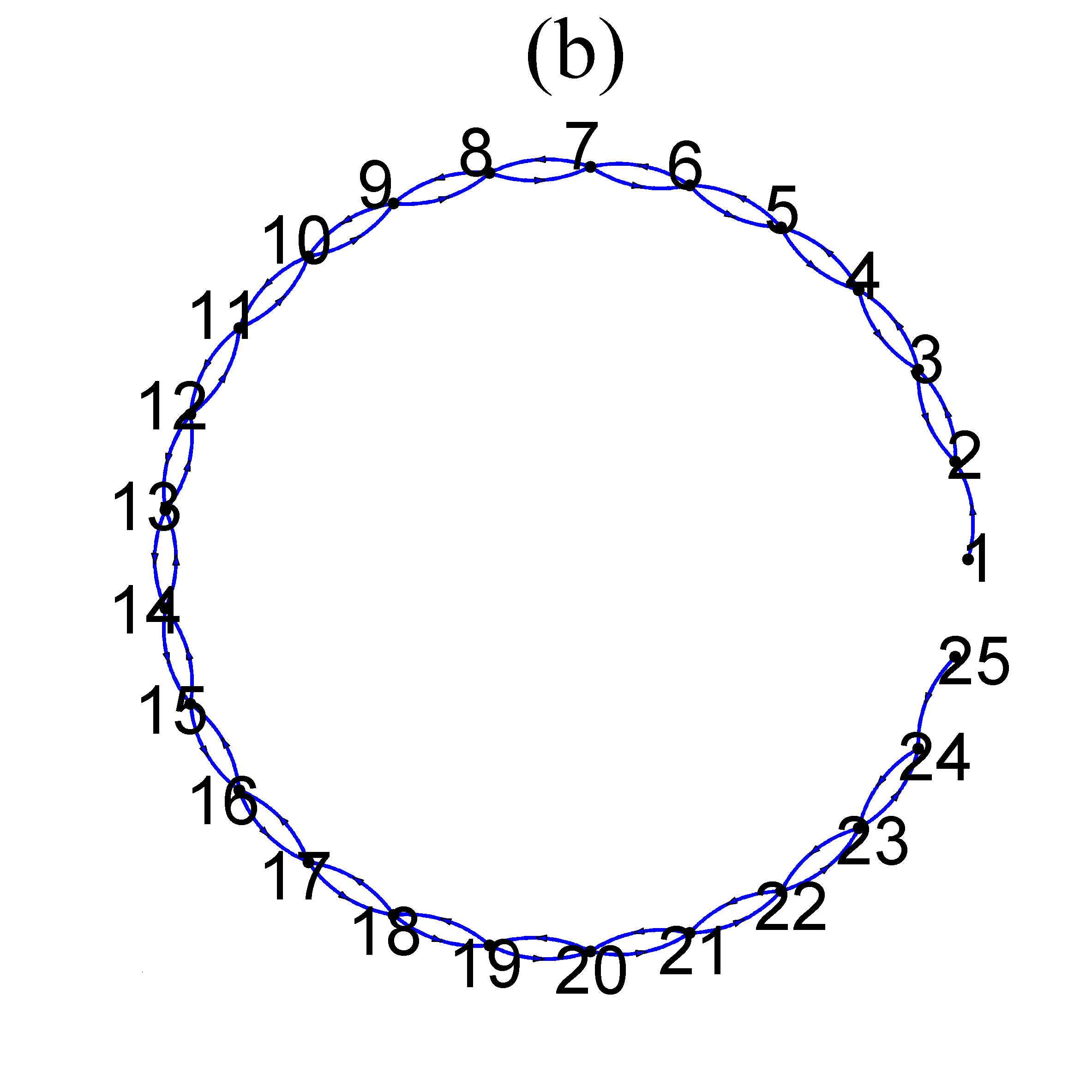}
\includegraphics[scale=0.25]{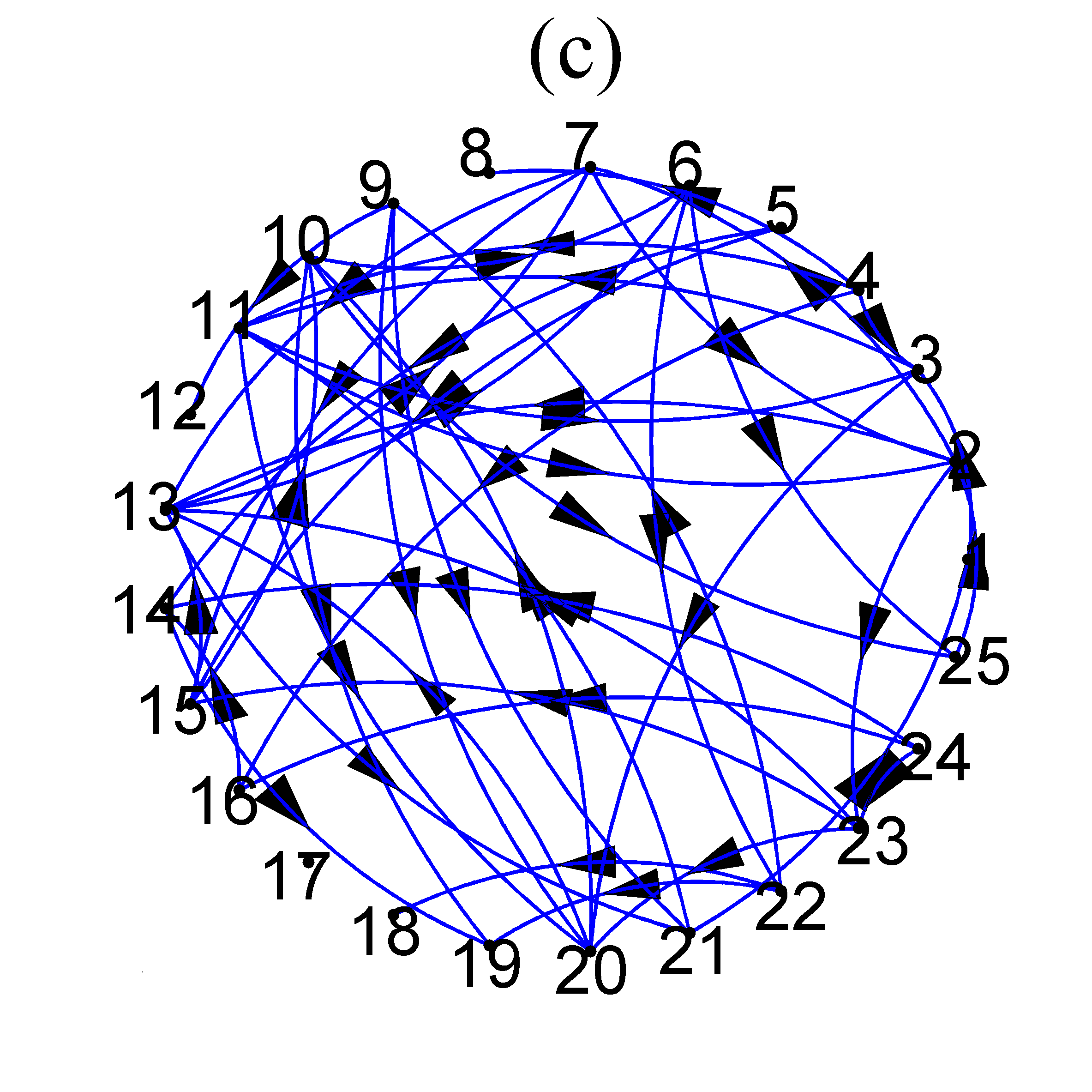}
\caption{Coupling networks for S1, S2, S3 in (\textbf{a}) for $K\!=\!5$, in (\textbf{b}) for $K\!=\!25$ , and in (\textbf{c}) for S4 and $K\!=\!25$.}
\label{fig:connections}
\end{figure}

\begin{figure}[H]
\centering
\includegraphics[scale=0.2]{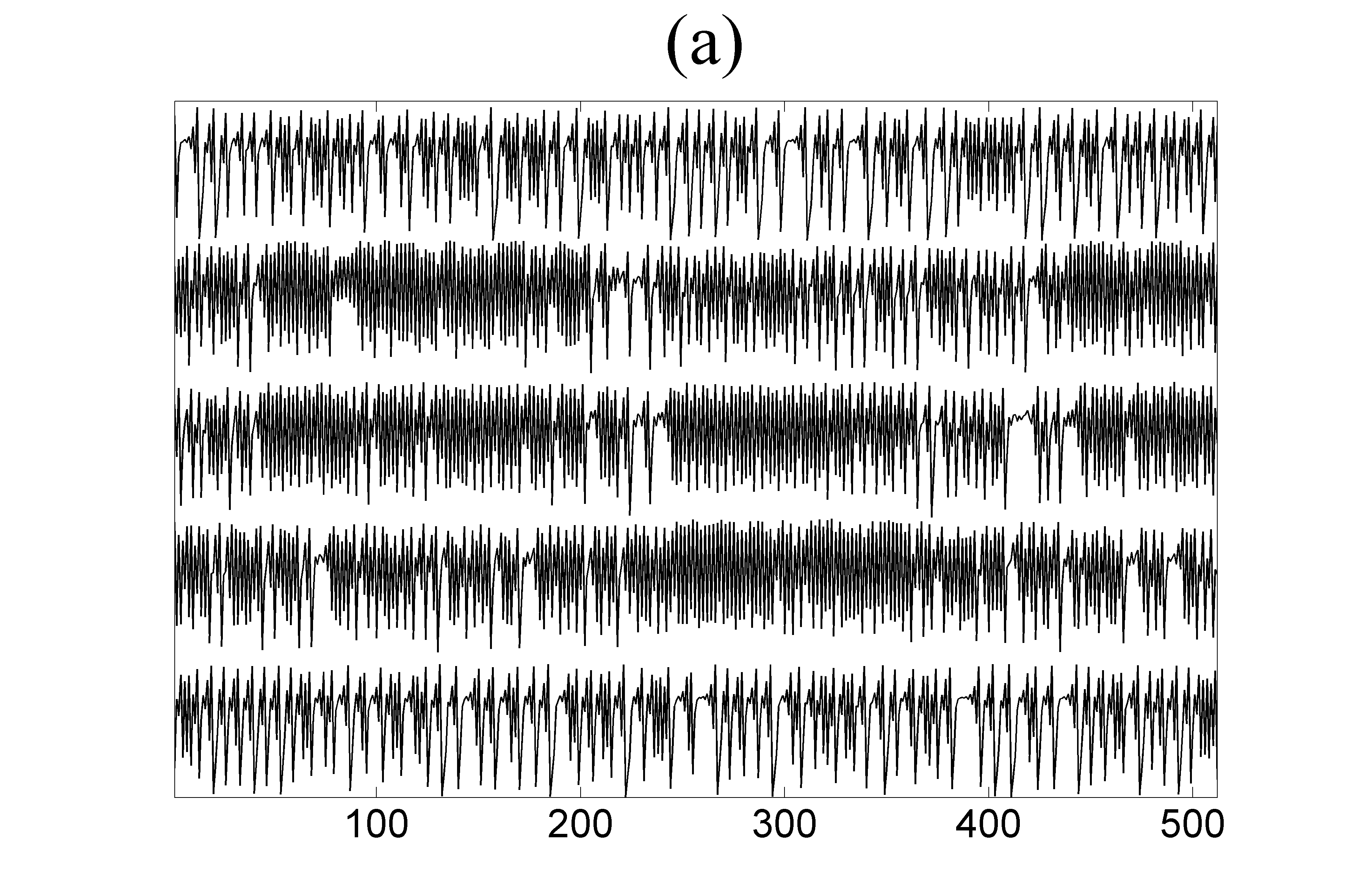}
\includegraphics[scale=0.2]{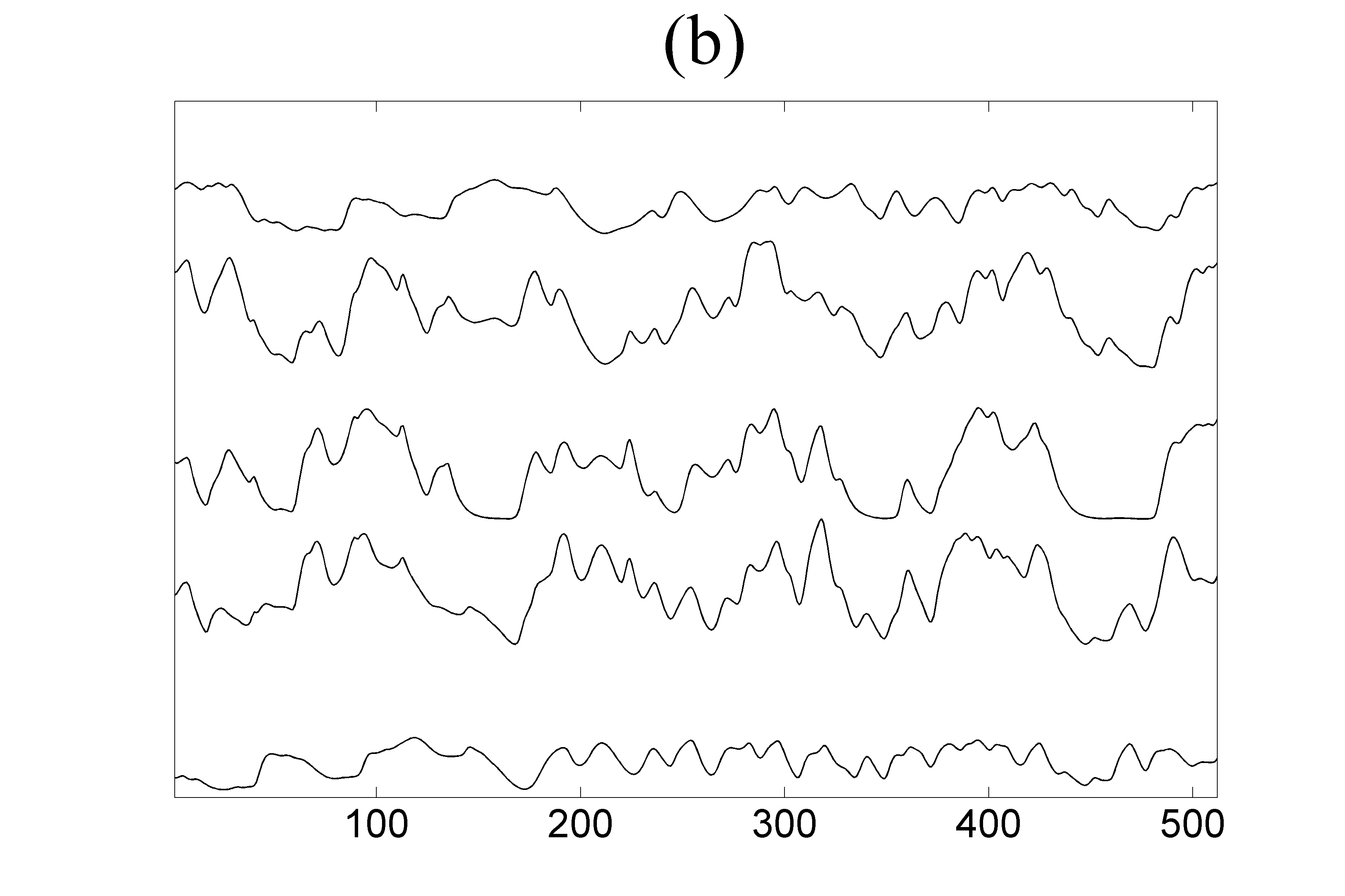}
\includegraphics[scale=0.2]{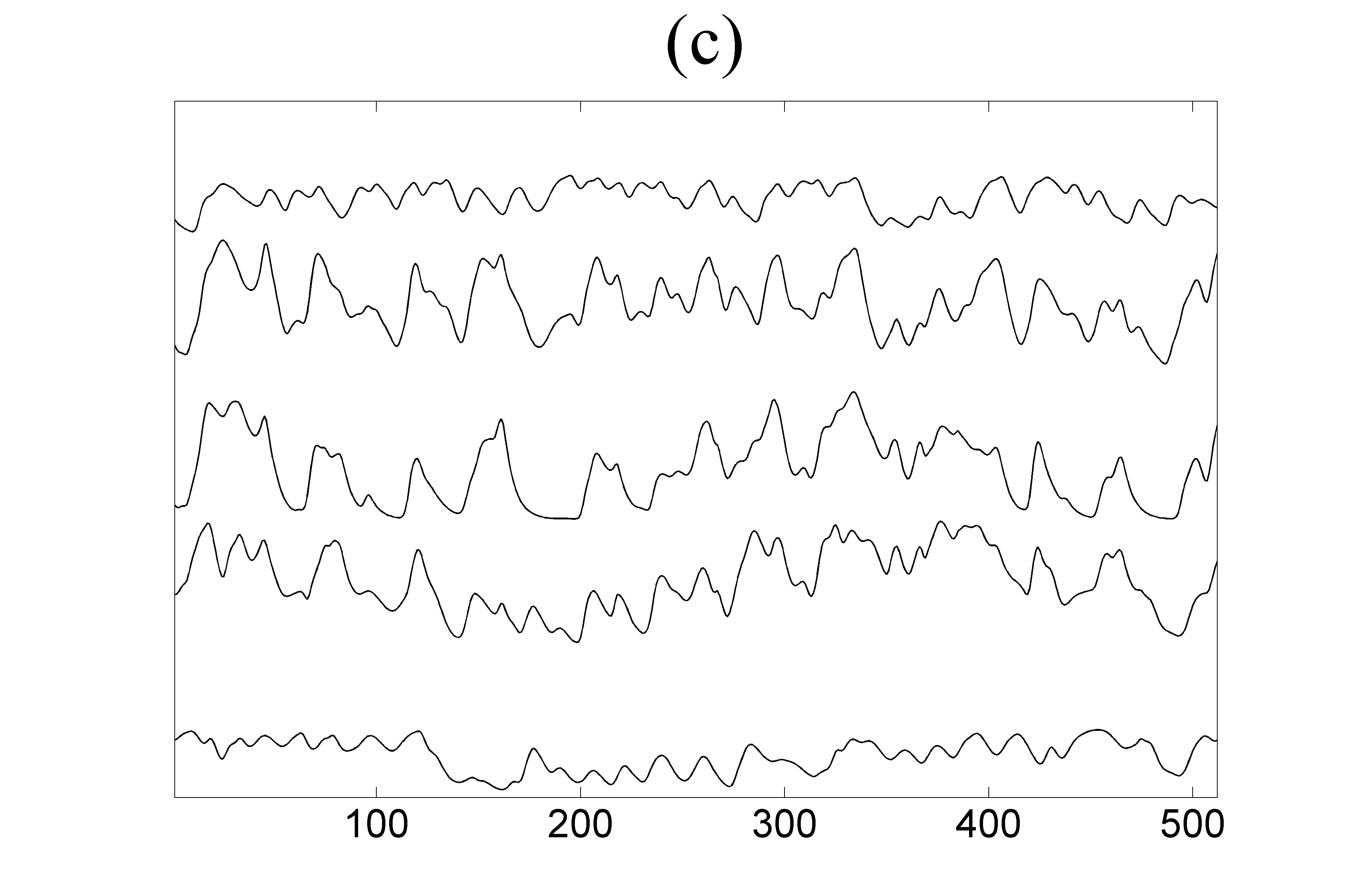}
\includegraphics[scale=0.2]{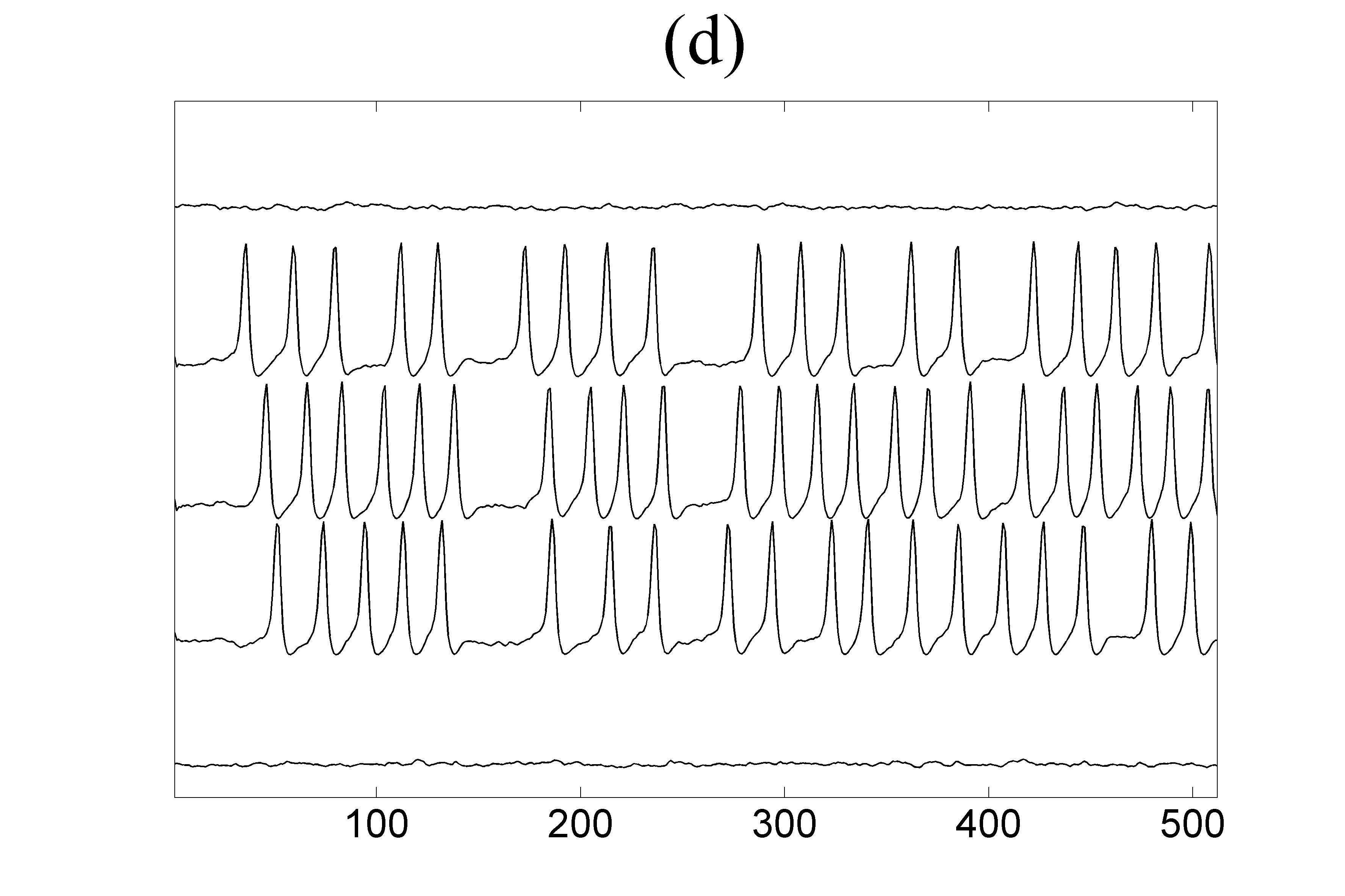}
\includegraphics[scale=0.2]{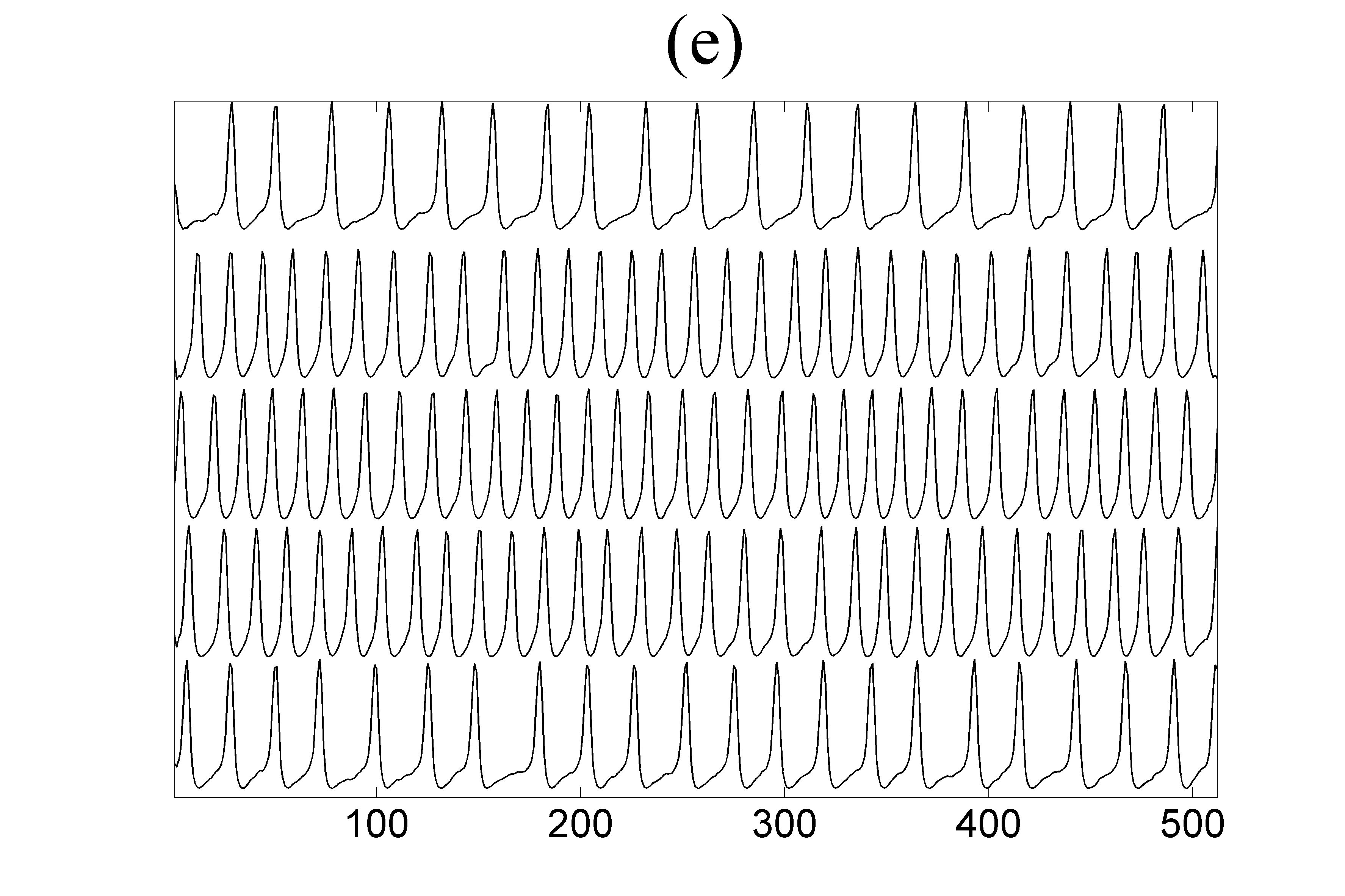}
\includegraphics[scale=0.2]{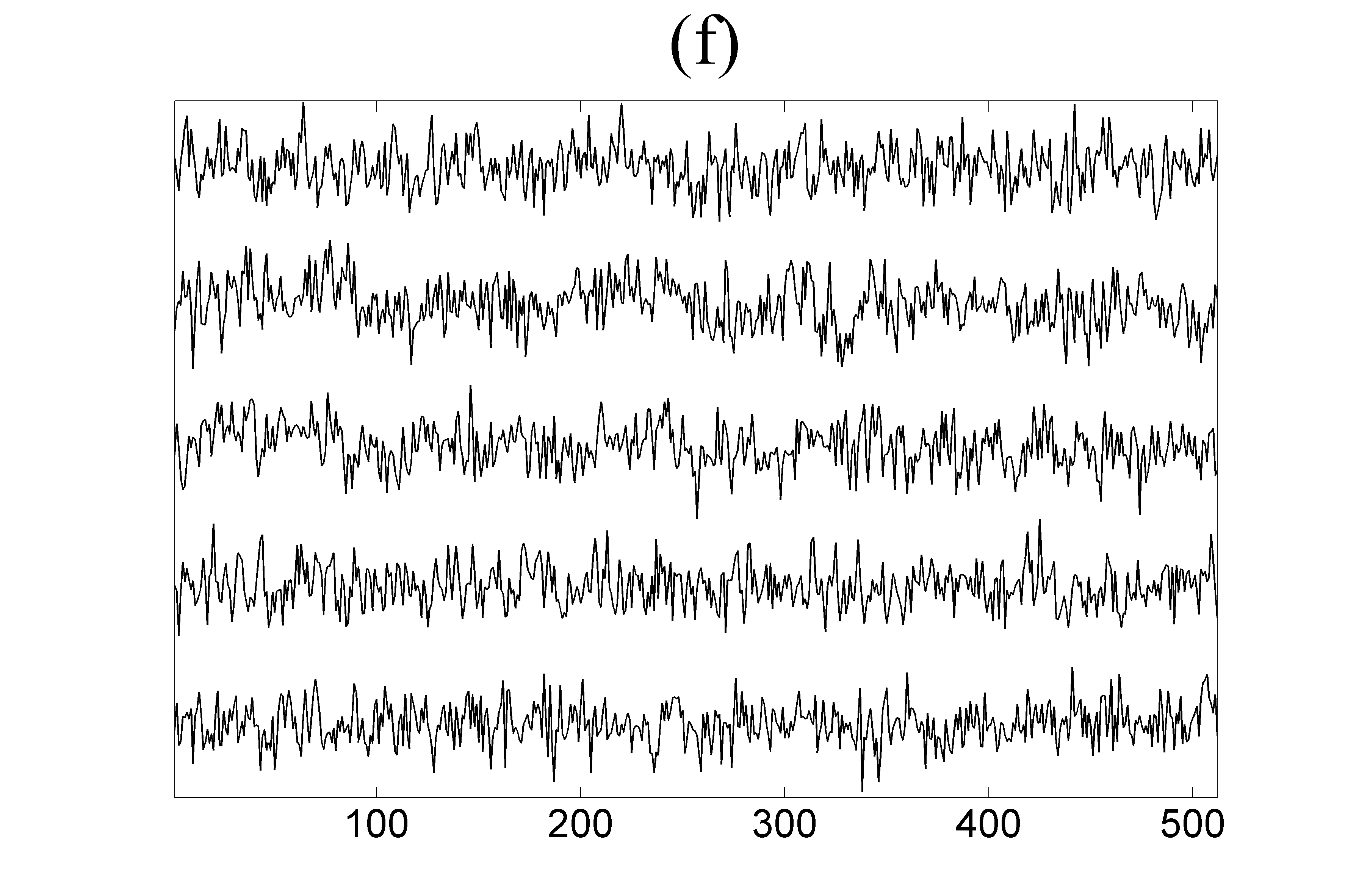}
\caption{Time series for: (\textbf{a}) coupled H\'{e}non maps for $C\!=\!0.2$,
    (\textbf{b}) coupled Mackey-Glass for $\Delta\!=\!100$ and $C\!=\!0.2$, (\textbf{c}) coupled Mackey-Glass
    for $\Delta\!=\!300$ and $C\!=\!0.2$, (\textbf{d}) neural mass for $A\!=\!3.45$ and $C\!=\!80$,
    (\textbf{e}) neural mass for $A\!=\!3.7$ and $C\!=\!80$, (\textbf{f}) VAR(3) model for $C\!=\!0.23$.}
\label{fig:timeseriessystems}
\end{figure}

{\bf S1}: The system of coupled H\'{e}non maps is a system of
coupled chaotic maps defined as
\begin{equation}
\hspace{-5mm}
 \begin{array}{ll}
x_{i,t} = 1.4-x_{i,t-1}^2+0.3x_{i,t-2}, & \mbox{for} \,\, i=1,K  \\
x_{i,t} = 1.4 - \left( 0.5 C (x_{i-1,t-1} + x_{i+1,t-1}) +
(1-C)x_{i,t-1} \right)^2 + 0.3 x_{i,t-2}, & \mbox{for} \,\,
j=2,\ldots,K-1
 \end{array}
\label{eq:Henon}
\end{equation}
where $K$ denotes the number of variables and $C$ is the coupling
strength. We consider the system for $K\!=\!5$ and $K\!=\!25$ and the
corresponding coupling network is shown in
Fig.~\ref{fig:connections}a and b, respectively.
Multivariate time series of size $K$ are generated and we use
short and long time series of length $n\!=\!512$ and $n\!=\!2048$,
respectively. An exemplary time series for $K\!=\!5$ is given in
Fig.~\ref{fig:timeseriessystems}a.

{\bf S2}: The system of coupled Mackey-Glass is a system of
coupled identical delayed differential equations defined as
 \begin{equation}
 \dot{x}_j(t) = -0.1x_j(t)+
 \sum_{i=1}^{K}\frac{C_{ij}x_i(t-\Delta)}{1+x_i(t-\Delta)^{10}}
 \quad\quad \mbox{for} \,\, j=1,2,\ldots,K \label{eq:MG}
 \end{equation}
where $K$ is the number of subsystems coupled to each other,
$C_{ij}$  is the coupling strength and $\Delta$ is the lag
parameter. 
We set $C_{ii}\!=\!0.2$ and $C_{i,j}$ for $i\neq j$ zero or $C$ 
according to the ring coupling structure shown in 
Fig.~\ref{fig:connections}a and b for $K\!=\!5$ and $K\!=\!25$, respectively.
For details on the solution of the delay
differential equations and the generation of the time series, see
\cite{Kugiumtzis13a}. Two scenarios are considered regarding the
inherent complexity of each of the $K$ subsystems given by $\Delta
\! = \! 100$ and $\Delta \! = \! 300$, regarding high complexity
(correlation dimension is about 7.0 \cite{Kugiumtzis96}) and even
higher complexity (not aware of any specific study for this
regime), respectively. Exemplary time series for each $\Delta$ and
$K\!=\!5$ are given in Fig.~\ref{fig:timeseriessystems}b and c. The
time series used in the study have length $n\!=\!4096$.

{\bf S3}: The neural mass model is a system of coupled
differential equations with a stochastic term that produces time
series similar to EEG simulating different states of brain
activity, e.g. normal and epileptic. It is defined as
\begin{equation}\label{eq:NMM}
\begin{aligned}
\dot{y}_0^j(t) =& y_3^j(t) \\
\dot{y}_3^j(t) =& AaS[y_1^j(t)-y_2^j(t)]-2ay_3^j(t)-a^2y_0^j(t) \\
\dot{y}_1^j(t) =& y_4^j(t) \\
\dot{y}_4^j(t) =& Aa\left\{p^j(t)+C_2S(C_1y_0^j)+\sum_{\substack{i=1 \\ i\neq j}}^K C_{ij} y_6^i(t)  \right\} \\
&-2ay_4^j(t)-a^2y_1^j(t) \\
\dot{y}_2^j(t) =& y_5^j(t) \\
\dot{y}_5^j(t) =& Bb\left\{C_4S[C_3y_0^j(t)]\right\}-2by_5^j(t)-b^2y_2^j(t) \\
\dot{y}_6^j(t) =& y_7^j(t) \\
\dot{y}_7^j(t) =&
Aa_dS(y_1^j(t)-y_2^j(t))-2a_dy_7^j(t)-a_d^2y_6^j(t)
\end{aligned}
\end{equation}
where $j$ denotes each of the $K$ subsystems representing the
neuron population defined by eight interacting variables and the
population (subsystem) interacts with other populations through
variable $y_4^j$ with coupling strength $C_{ij}$. We set
$C_{ii}\!=\!0.0$ and $C_{i,j}$ for $i\neq j$ zero or $C$ according to the ring coupling structure
shown in Fig.~\ref{fig:connections}a and b for $K\!=\!5$ and $K\!=\!25$,
respectively. The term $p^j(t)$ represents a random influence from
neighboring or distant populations, $A$ is an excitation parameter
and $B$, $a$, $b$, $a_d$, $C1$-$C4$ other parameters (see
\cite{Wendling00} for more details). The function $S$ is the
sigmoid function $S(v)\!=\!2e_0/(1+e^{r(v_0-v)})$, where $r$ is the
steepness of the sigmoid and $e_0$, $v_0$ are other parameters
explained in \cite{Wendling00}. From each population $j=1,\ldots,K$ we
consider only the first variable ${y}_0^j$ and obtain the
multivariate time series of $K$ variables. The value of the
excitation parameter $A$, affects the form of the output signals
combined with the coupling strength level, ranging from similar to
normal brain activity with no spikes to almost periodic with many
spikes similar to epileptic brain activity. We consider two values
for this parameter, one for low excitation with $A\!=\!3.45$ and one
for high excitation with $A\!=\!3.7$. Exemplary time series for each
$A$ and $K\!=\!5$ are given in Fig.~\ref{fig:timeseriessystems}d and
e. The time series used in the study have length $n\!=\!4096$.

{\bf S4}: The VAR process on $K\!=\!25$ variables and order $P\!=\!3$ as
suggested in \cite{Basu15} is used as representative of a
high-dimensional linear stochastic process. Initially, 4\% of the
coefficients (total coefficients 1875) of VAR(3) selected randomly
are set to 0.9 and the rest are zero and the positive coefficients
are reduced iteratively until the stationarity condition is
fulfilled. The autoregressive terms of lag one are set to one. The
true couplings are 8\% of a total of 600 possible ordered
couplings. An exemplary coupling network of random type is shown
in Figure \ref{fig:connections}(c). The time series length is set
to $n\!=\!512$ and an exemplary time series of only five of the $K\!=\!25$
variables of the VAR(3) process is shown in
Fig.~\ref{fig:timeseriessystems}f.

For S1, S2 and S3, the coupling strength $C$, fixed for all
couplings, is varied to study a wide range of coupling levels from
zero coupling to very strong coupling. Specifically, for S1 and S2
$C\!=\!0,0.05,0.1,0.2,0.3,0.4,0.5$, and for S3
$C\!=\!0,20,40,80,120,160,200$. For S4, only one case of coupling
strength is considered, given by the magnitude $0.23$ of all
non-zero coefficients, for which the stationarity of VAR(3)
process is reached. For each system and scenario of different
coupling strength, 10 multivariate time series (realizations) are generated to
obtain statistically valid results. The evaluation is performed as described in Section~\ref{subsec:score}.
 
\section{Results}
\label{sec:Results}

In this section, the evaluation of the performance of all
causality measures is presented for each system and scenario.
First, the procedure of the evaluation is shown in one specific
setting, then the measures are evaluated and ranked for each
system and finally the overall ranking is given.

\subsection{Evaluation of measures in one exemplary setting}

\begin{figure}[H]
\centering
\includegraphics[scale=0.25]{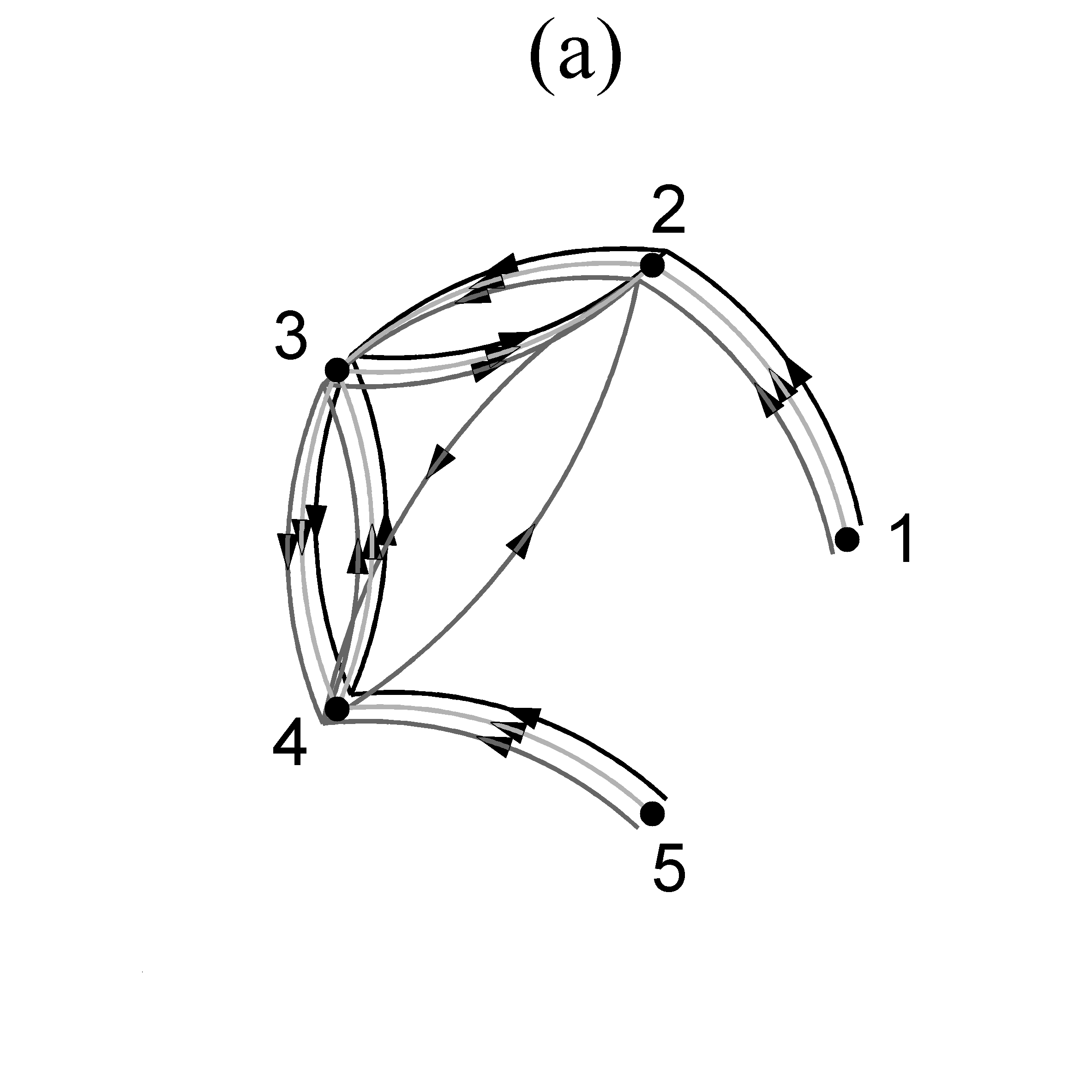}
\includegraphics[scale=0.25]{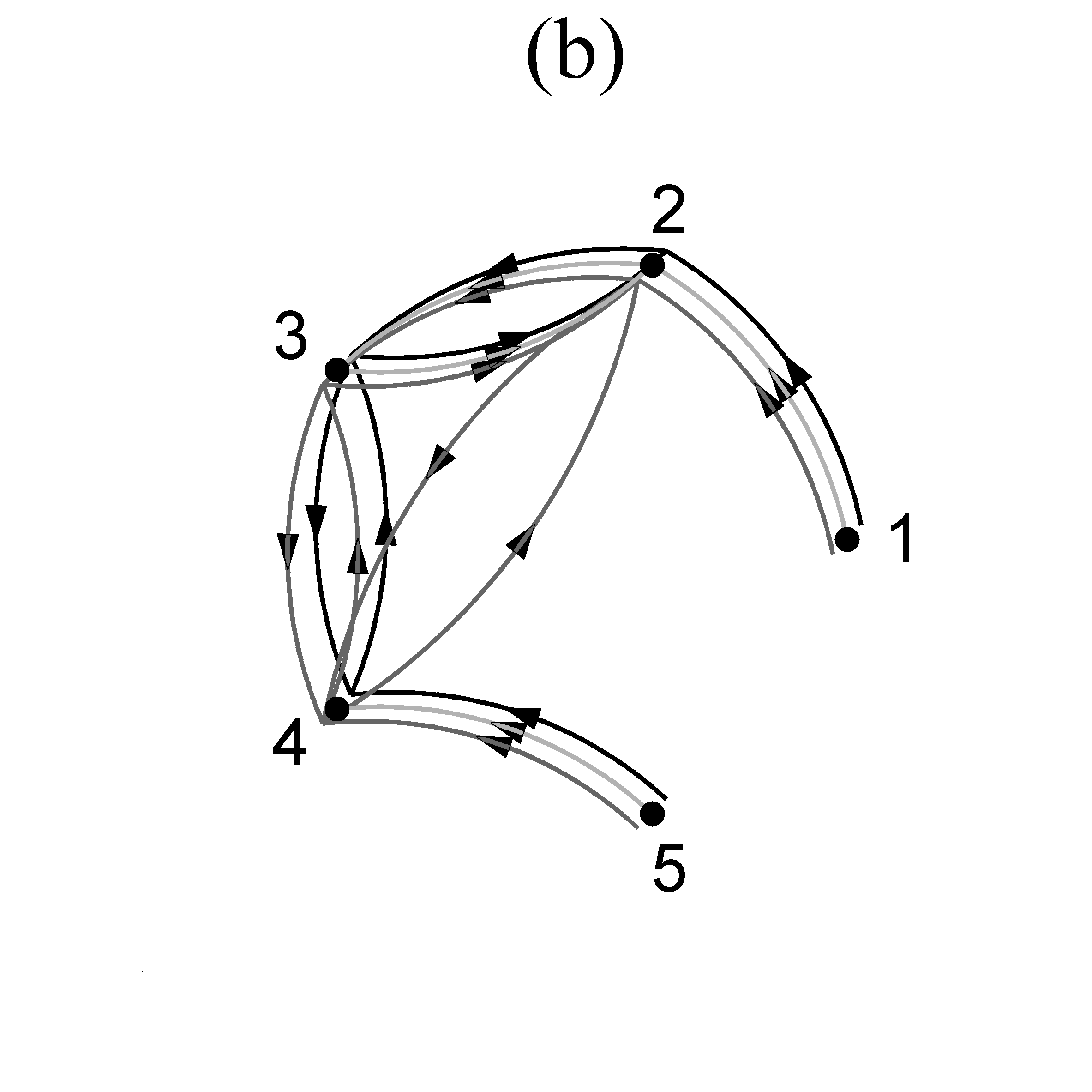}
\includegraphics[scale=0.25]{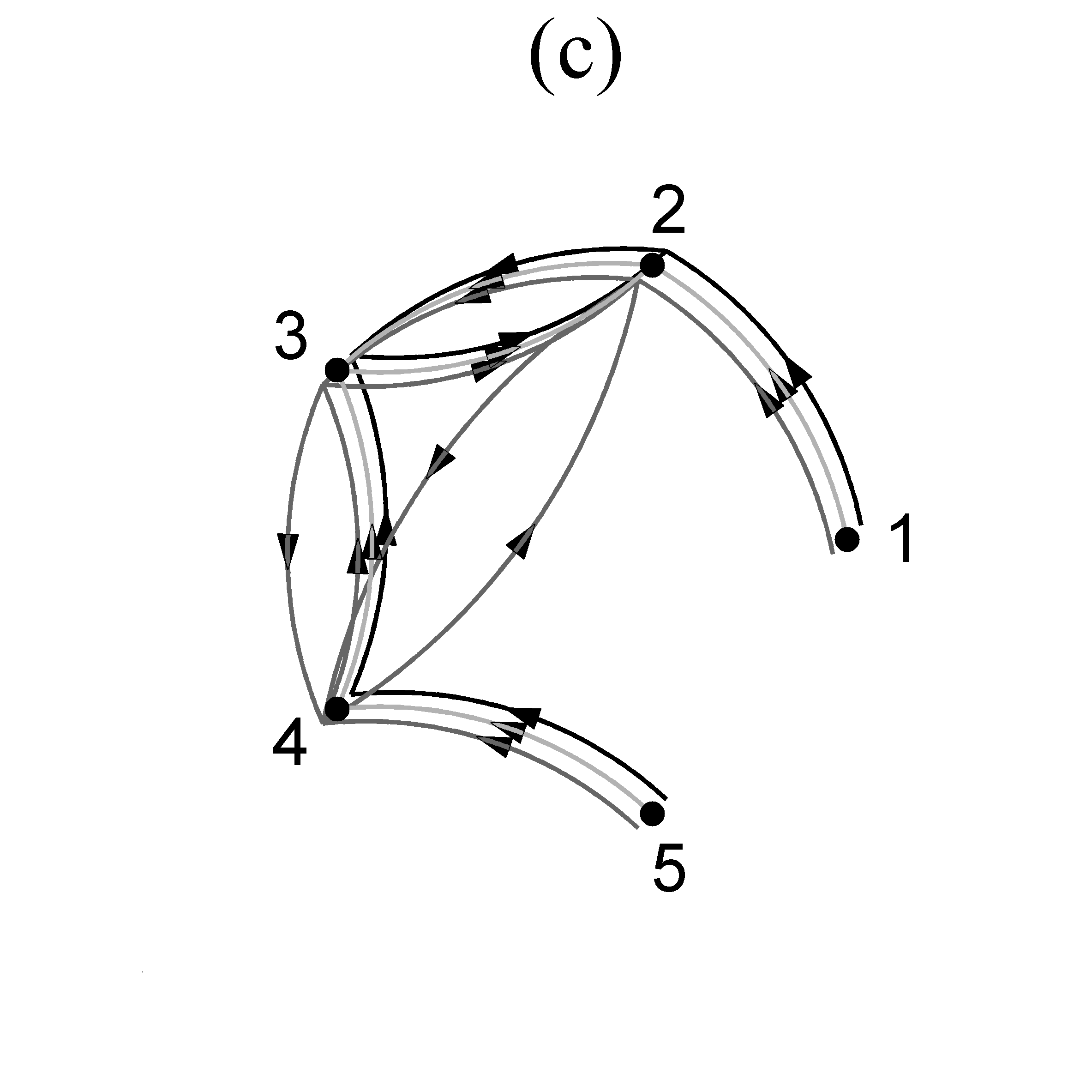}
\caption{Binary networks of TE measure for S1 using
    (\textbf{a}) statistical significance test (light grey $\rightarrow$ $\alpha\!=\!0.01$ ,
    black $\rightarrow$ $\alpha\!=\!0.05$, grey $\rightarrow$ $\alpha\!=\!0.1$),
    (\textbf{b}) density threshold (light grey $\rightarrow$ $\rho\!=\!4$ , black $\rightarrow$ $\rho\!=\!6$, grey $\rightarrow$ $\rho\!=\!8$)
    and  (\textbf{c}) magnitude threshold (light grey $\rightarrow$ $th_{4}$ , black $\rightarrow$ $th_{6}$, grey $\rightarrow$
    $th_{8}$)}
    \label{fig:surdensthres}
\end{figure}

We consider a multivariate time series of length $n\!=\!512$ from the
system S1 of coupled H\'{e}non maps for $K\!=\!5$ variables and
coupling strength $C\!=\!0.2$. The original coupling network has the
ring structure as shown in Fig.~\ref{fig:connections}a. We derive
the estimated causality (weight) matrix by the bivariate measure
of transfer entropy (TE) using the appropriate parameters of
embedding dimension $m\!=\!2$ and $\tau\!=\!1$
\begin{equation}
R_{\mbox{\scriptsize TE}}=\left( \begin{array}{ccccc}
  0 &  \textbf{0.148} & -0.003 & -0.002 & -0.012 \\
  -0.005 &  0 & \textbf{0.167} & \textbf{0.013} & -0.021 \\
  -0.015 &  \textbf{0.094} & 0 & \textbf{0.051} & -0.004 \\
  -0.015 &  \textbf{0.016} & \textbf{0.092} & 0 & -0.014 \\
  -0.011 &  -0.009 & 0.003 & \textbf{0.193} & 0 \\
\end{array} \right)
\label{fig:TEone}
\end{equation}
where the negative values denote the negative bias in the
estimation of TE with the nearest neighbors estimate. Applying the
three criteria of measure significance for transforming the weight
matrix to an adjacency matrix (see Sec.~\ref{subsec:Network}), we
derive the causality binary networks. Specifically, as shown in
Fig.~\ref{fig:surdensthres}, different binary networks are
obtained for the different values of the significance level of the
randomization test, the network density and the magnitude
threshold.

In Fig.~\ref{fig:surdensthres}a, the binary network for
significance level $\alpha\!=\!0.01$ and $\alpha\!=\!0.05$ coincides with
the original coupling network, whereas for $\alpha\!=\!0.1$ more
connections are present regarding indirect causal effects (for the
latter the statistical significant values are given in bold in the
form of eq.(\ref{fig:TEone})). When preserving the 4, 6, and 8
strongest connections, as shown in Fig.~\ref{fig:surdensthres}b,
the true structure is preserved only when the correct density is
set (6 connections), indicating that the highest TE values are
reached at the true couplings. This seems the optimal strategy for
thresholding, but in real-world applications the actual network
density is not a-priori known. Similarly, in
Fig.~\ref{fig:surdensthres}c, the binary networks obtained using
three magnitude thresholds on TE values are shown. Each of the three
magnitude thresholds is computed as the average threshold for preserving the corresponding network density across 10 realizations, for this specific coupling strength and causality measure.
These magnitude thresholds happen not to be the ones
corresponding to the network densities for this realization and
actually none of the three thresholds identifies all the existing
and non-existing connections.

For illustration purposes, we compute the performance indices for
TE at this scenario using the statistical significance criterion
for $\alpha\!=\!0.1$, given in Table~\ref{tab:TEperformance}.
\begin{table}[H]
\caption{Computation of the performance indices for the causality
measure TE, where the binary causality network is derived using
the statistical significance criterion for $\alpha\!=\!0.1$.}
\centering
\begin{tabular}{lcc}
\toprule
              & \textbf{true positive} & \textbf{true negative} \\
\textbf{positive found} & $\mbox{TP} = 6$        & $\mbox{FP} = 2$        \\
\textbf{negative found} & $\mbox{FN} = 0$        & $\mbox{TN} = 12$   \\
\midrule
\multicolumn{3}{l}{sens : $6/6=1$}          \\
\multicolumn{3}{l}{spec : $12/14=0.86$}         \\
\multicolumn{3}{l}{MCC : {\Large $\frac{6 \cdot 12-2 \cdot
0}{\sqrt{(6+2) \cdot (6+0) \cdot (12+2) \cdot (12+0)}}$}$=0.80$}\\
\multicolumn{3}{l}{FM : {\Large $\frac{2 \cdot 6}{2 \cdot 6 + 0 + 2}$}$=0.86$}\\
\multicolumn{3}{l}{HD : $2+0=2$}\\
\bottomrule
\end{tabular}
 \label{tab:TEperformance}
\end{table}
We note that the two extra connections found significant using
$\alpha\!=\!0.1$ reduce the specificity to 0.86 while sensitivity is
one, which affects accordingly the other three measures. Note that
the mismatch of just two out of 20 connections (HD=2) gives
MCC=0.8, significantly lower from one, and the same holds for the
F-measure index.

For the same scenario, the measure of PMIME (for a maximum lag
$L\!=\!5$ well above the optimal lag 2) gives a weight matrix of
zero and positive numbers
\[R_{\mbox{\scriptsize PMIME}}=\left(
\begin{array}{ccccc}
  0 &  0.162  & 0      & 0      & 0 \\
  0 &  0       & 0.101 & 0      & 0 \\
  0 &  0.076  & 0      & 0.077 & 0 \\
  0 &  0       & 0.094 & 0      & 0 \\
  0 &  0       & 0      & 0.173 & 0 \\
\end{array} \right)\]
No significance criterion is applied here, and simply setting the
positive numbers to one gives the adjacency matrix, and in this
case the estimated causality network matches exactly the original
coupling network giving HD=0 and all other indices equal to one.

\subsection{Results with respect to performance indices,
significance criteria and coupling strength}

First, we give a comprehensive presentation reporting all the
performance indices presented in Sec.~\ref{subsec:accuracy} and
the significance criteria in Sec.~\ref{subsec:Network} for system
S1 and $K\!=\!25$, $n\!=\!2048$ and $C\!=\!0.2$. In Table~\ref{tab:RANKHMMCC},
the five performance indices of the eight highest ranked causality
measures in terms of MCC are presented.
\begin{table}[H]
\caption{The rankings of the eight best measures according to MCC
for coupled H\'{e}non maps ($K\!=\!25$, $n\!=\!2048$, $C\!=\!0.2$) are
presented for the three binarization methods. The sensitivity
(sens), specificity (spec), F measure (FM) and Hamming distance
(HD) performance indices are also presented.}
\centering
\begin{tabular}{llccccc}
\toprule
& Measure      & sens & spec & MCC & FM & HD \\
\midrule
\multicolumn{7}{c}{statistical significance test ($\alpha\!=\!0.05$)}\\
\midrule
1 & \textbf{PMIME}($L\!=\!5$) &0.79 & 0.99 & 0.86 & 0.86 & 11\\
2 & \textbf{RGPDC}($p\!=\!5$,$\beta$) & 0.86 & 0.85 & 0.49 & 0.49   & 84.5\\
3 & \textbf{RGPDC}($p\!=\!5$,$\alpha$)    & 0.87      & 0.84& 0.47     & 0.47    & 91.6\\
4 & GPDC($p\!=\!5$,$\beta$)     & 0.87      & 0.84 & 0.47     & 0.46   &92.3         \\
5 & \textbf{RCGCI}($p\!=\!5$) & 0.92      & 0.81      &0.47 & 0.45   & 106.1\\
6 & \textbf{RGPDC}($p\!=\!5$,$\gamma$)    & 0.86       & 0.83& 0.46     & 0.46   & 95.8\\
7 & GCI($p\!=\!5$)        & 0.83      & 0.84& 0.45     & 0.45   & 92.9          \\
8 & \textbf{RGPDC}($p\!=\!5$,$\theta$)& 0.83      & 0.84      & 0.45 & 0.45&95.4\\
\midrule
\multicolumn{7}{c}{density threshold ($\rho_0\!=\!48$)}\\
\midrule
1 & \textbf{PMIME}($L\!=\!5$) & 0.79 & 0.99      & 0.86&0.86   & 10.9\\
2 & TE($m\!=\!2$)        & 0.68      & 0.97& 0.66     & 0.68   & 28.8          \\
3 & TE($m\!=\!3$) & 0.67      & 0.97      & 0.64 & 0.67   &30.2\\
4 & PGCI($p\!=\!5$)       & 0.6           & 0.96& 0.56    & 0.6        & 36.8\\
5 & GPDC($p\!=\!5$,$\beta$)     & 0.59      & 0.96& 0.56     & 0.59   & 37            \\
6 & GPDC($p\!=\!5$,$\alpha$)     & 0.57      & 0.96      & 0.54 & 0.57 & 38.8\\
7 & \textbf{RGPDC}($p\!=\!5$,$\beta$)  & 0.56 & 0.96      &0.52     & 0.56   & 40\\
8 & CGCI($p\!=\!5$)    & 0.56      & 0.96 & 0.52     & 0.56   & 40            \\
\midrule
\multicolumn{7}{c}{magnitude threshold ($th_{48}$)}\\
\midrule
1 & \textbf{PMIME}($L\!=\!5$) & 0.78      & 0.99     & 0.86 &0.86 & 11\\
2 & TE($m\!=\!2$)        & 0.67 & 0.97 & 0.65     & 0.67   & 29.9\\
3 & TE($m\!=\!3$)      & 0.66       & 0.97 & 0.64     & 0.67 &29.9          \\
4 & GPDC($p\!=\!5$,$\beta$)    & 0.58      & 0.96 &0.55 & 0.57 & 39.8\\
5 & GPDC($p\!=\!5$,$\alpha$)     & 0.57 & 0.95      &0.53     & 0.56  & 41.8\\
6 & \textbf{RGPDC}($p\!=\!5$,$\beta$)  & 0.54      & 0.96& 0.51     & 0.55   & 40.4          \\
7 & PGCI($p\!=\!5$)    & 0.50 & 0.96      & 0.51     & 0.52  &40.6\\
8 & \textbf{RGPDC}($p\!=\!5$,$\alpha$)       & 0.53 & 0.96      & 0.50     & 0.54  &41.9\\
\bottomrule
\end{tabular}
\label{tab:RANKHMMCC}
\end{table}
In Table~\ref{tab:RANKHMMCC} and all tables to follow, the
measures making use of dimension reduction, i.e. PMIME, RCGCI and
RGPDC, are highlighted (bold face) to accommodate comparison with
the other measures. The results are organized in three blocks, one
for each of the three significance criteria. For the criterion of
statistical significance at $\alpha\!=\!0.05$, the dimension reduction
measures score highest in all performance indices. The PMIME
($L\!=\!5$) measure obtains the greatest specificity value $0.99$ and
RCGCI ($p\!=\!5$) the greatest sensitivity value $0.92$. A large
difference between the first MCC=0.86 for PMIME and the MCC for
the other highest ranked measures is observed, while for the
specificity and sensitivity indices this does not hold. This is
explained by the fact that a small decrease in specificity implies
increase in the number of falsely detected causal effects that for
networks of low density dominates in the determination of MCC (see
eq.(\ref{eq:MCCdef})). For the significance criteria of density
($\rho_0\!=\!48$ equal to the number of true couplings) and threshold
($th_{48}$), PMIME ($L\!=\!5$) is unaltered at first rank while the
information and frequency measures exhibit better performance
compared to the criterion of statistical significance. It is also
observed that these two criteria show lower sensitivity and higher
specificity, which questions the rule that the couplings of
largest causality values are the true ones. When $\rho$ is smaller
or larger than the true density, sensitivity changes more than
specificity again due to the sparseness of the true network.
Similar conclusions are inferred for the significance criterion of
magnitude threshold.

We demonstrate further the dependence of the causality measure
performance on the parameter in each significance criterion using
the S4 system of VAR process ($K\!=\!25$, $n\!=\!512$, $P\!=\!3$). In
Table~\ref{tab:VARMCC}, the ranking of the eight best measures in
terms of MCC is presented for three parameters of each of the
three significance criteria.
\begin{table}[H]
\caption{The rankings of the eight best measures according to MCC
for the system S4 of the VAR process ($K\!=\!25$, $n\!=\!512$, $p\!=\!3$) in
conjunction with each significance criterion and its parameter.
Three rankings are given in three blocks, one for each
significance criterion and for three different choices of its
parameter, where $\rho_0\!=\!48$  is the true density.}
\centering
\begin{tabular}{llccc}
\toprule
\multicolumn{5}{c}{statistical significance test}\\
\midrule
  & Measure      & $a\!=\!0.01$ & $a\!=\!0.05$ & $a\!=\!0.1$  \\
\midrule
1 & \textbf{RGPDC($p\!=\!3$, $\alpha$)} & 0.944 & 0.868 & 0.861\\
2 & \textbf{RGPDC($p\!=\!5$, $\alpha$)} & 0.944 & 0.867 & 0.861\\
3 & \textbf{RGPDC($p\!=\!3$, $\theta$)} & 0.940 & 0.868 & 0.861\\
4 & \textbf{RGPDC($p\!=\!5$, $\theta$)} & 0.939 & 0.867 & 0.861  \\
5 & \textbf{RGPDC($p\!=\!3$, $\delta$)} & 0.938 & 0.868 & 0.861\\
6 & \textbf{RGPDC($p\!=\!5$, $\delta$)} & 0.936 & 0.867 & 0.861\\
7 & \textbf{RGPDC($p\!=\!5$, $\beta$)}  & 0.933 & 0.867 & 0.861          \\
8 & \textbf{RCGCI($p\!=\!3$)}       & 0.933 & 0.867 & 0.861\\
\midrule
\multicolumn{5}{c}{density threshold}\\
\midrule
  & Measure      & 0.6$\rho_0$    & $\rho_0$    & 1.4$\rho_0$  \\
\midrule
1 & \textbf{RCGCI($p\!=\!3$)}       & 0.758 & 0.974 & 0.862\\
2 & \textbf{RGPDC($p\!=\!5$, $\alpha$)} & 0.758 & 0.972 & 0.862\\
3 & \textbf{RCGCI($p\!=\!5$)}       & 0.758 & 0.972 & 0.862\\
4 & \textbf{RGPDC($p\!=\!3$, $\alpha$)} & 0.758 & 0.972 & 0.862 \\
5 & \textbf{RGPDC($p\!=\!3$, $\gamma$)} & 0.755 & 0.972 & 0.862\\
6 & \textbf{RGPDC($p\!=\!5$, $\gamma$)} & 0.755 & 0.972 & 0.862\\
7 & \textbf{RGPDC($p\!=\!5$, $\theta$)} & 0.785 & 0.972 & 0.862\\
8 & \textbf{RGPDC($p\!=\!3$, $\beta$)} & 0.758 & 0.969 & 0.862   \\
\midrule
\multicolumn{5}{c}{magnitude threshold}\\
\midrule
  & Measure      & $th_{0.6\rho_0}$   & $th_{\rho_0}$   & $th_{1.4\rho_0}$  \\
\midrule
1 & \textbf{RCGCI($p\!=\!3$)}       & 0.751 & 0.979 & 0.868 \\
2 & \textbf{RGPDC($p\!=\!3$, $\theta$)} & 0.753 & 0.976 & 0.868\\
3 & \textbf{RGPDC($p\!=\!5$, $\theta$)} & 0.753 & 0.975 & 0.869\\
4 & \textbf{RGPDC($p\!=\!3$, $\alpha$)} & 0.757 & 0.975 & 0.868\\
5 & \textbf{RGPDC($p\!=\!3$, $\delta$)} & 0.750 & 0.975 & 0.868\\
6 & \textbf{RCGCI($p\!=\!5$)}       & 0.752 & 0.974 & 0.868 \\
7 & \textbf{RGPDC($p\!=\!5$, $\alpha$)} & 0.752 & 0.974 & 0.869\\
8 & \textbf{RGPDC($p\!=\!5$, $\delta$)} & 0.751 & 0.974 & 0.868\\
\bottomrule
\end{tabular}
\label{tab:VARMCC}
\end{table} 
For this linear system, the highest ranked causality measures for
all three significance criteria are the linear measures using
dimension reduction RCGCI and RGPDC for various parameter values.
This is somehow expected as these measures are both linear 
and the underlying system is linear, and they use dimension reduction as
the number of variables is $K=25$. For such a high-dimensional time series,
the bivariate linear measures give indirect (and false in our evaluation)
causality effects, whereas the multivariate linear measures without 
dimension reduction cannot reach the performance of RCGCI and RGPDC as the
time series length $n=512$ is relatively small for estimating accurately
the VAR model parameters (75 coefficients in VAR($3$) are to be estimated 
for each of the $K=25$ variables).
The best performance of the measures is achieved for the criterion
of statistical significance when $a\!=\!0.01$ and for the other two
significance criteria when the parameters corresponding to the
true density $\rho_0$, as expected. Comparing the three rankings
for the best parameter choice of each criterion, it is observed
that the statistical significance gives MCC values almost as high
as the other two methods. This fact indicates the advantage of the
statistical significance criterion, where the a-priori knowledge
of the true network density is not required. From this point on,
all the presented results are for the criterion of statistical
significance.

\begin{figure}[H]
\centering
\includegraphics[scale=0.3]{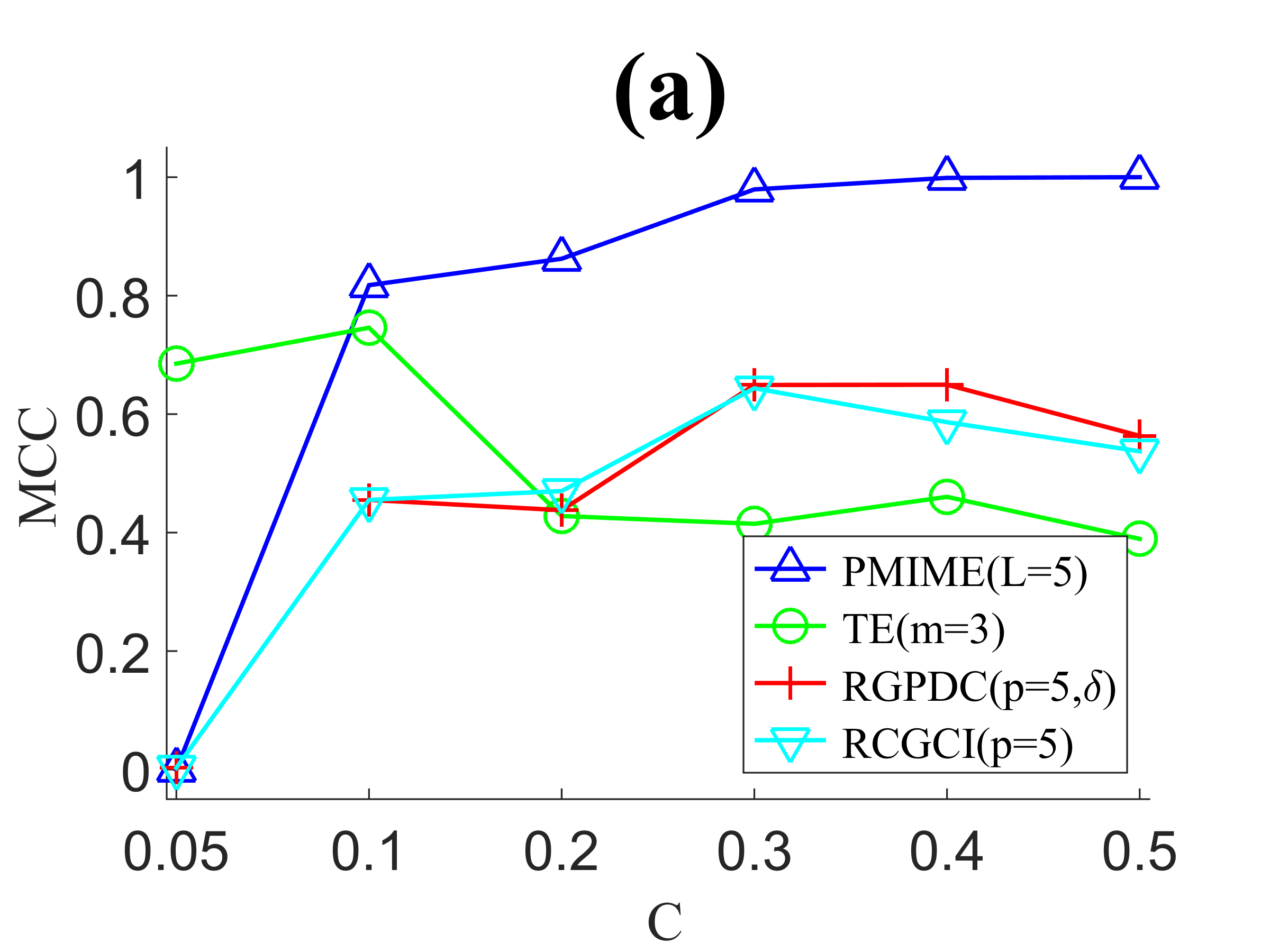}
\includegraphics[scale=0.3]{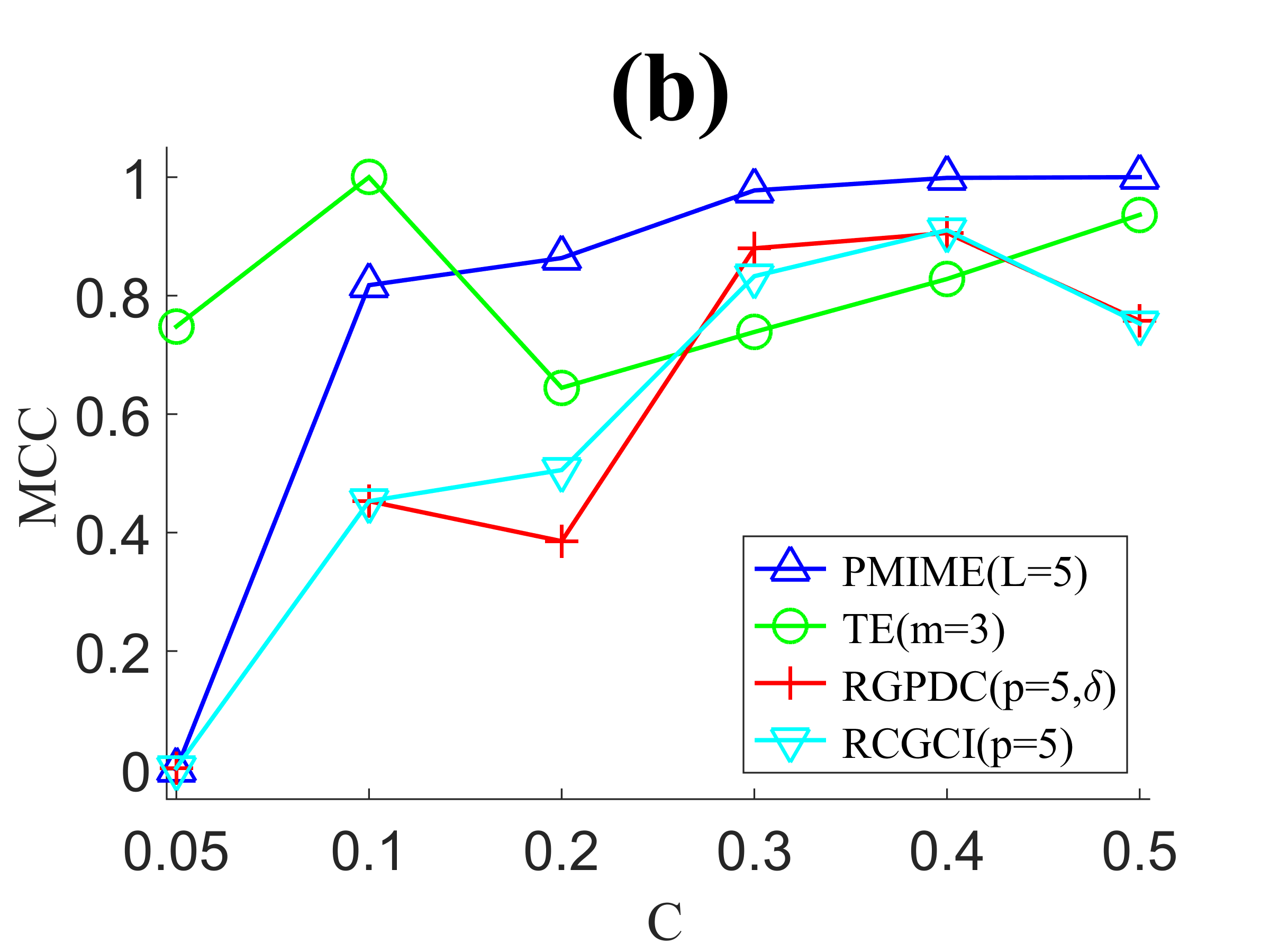}
\includegraphics[scale=0.32]{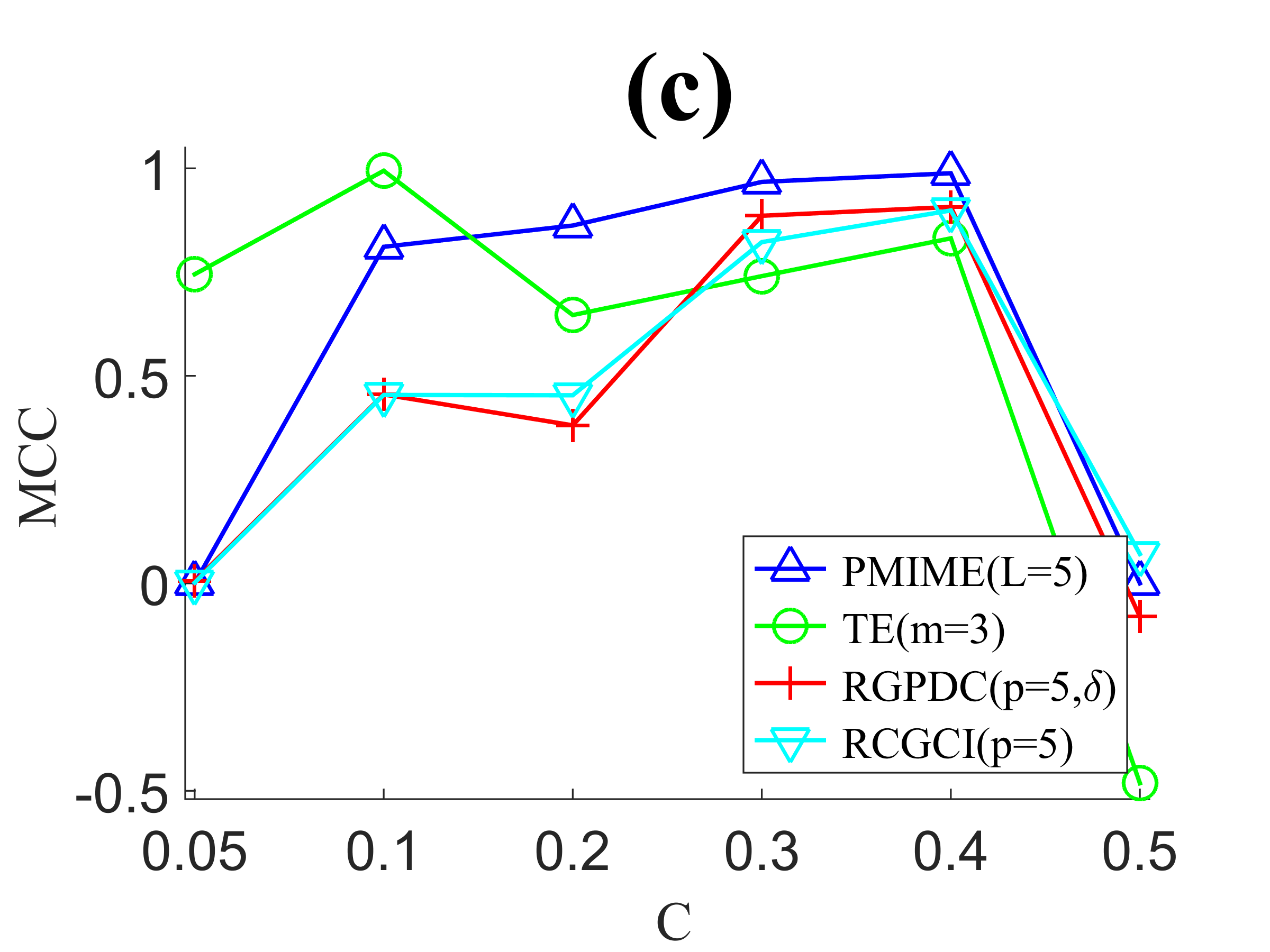}
\caption{MCC of PMIME ($L\!=\!5$), TE ($m\!=\!3$),
    RGPDC ($p\!=\!5$,$\delta$), RCGCI ($p\!=\!5$) as a function of the
    coupling strength $C$ in system S1 of the coupled H\'{e}non maps
    ($K\!=\!25$, $n\!=\!2048$) for the three significance criteria: (a) statistical testing
    at $\alpha\!=\!0.05$, (b) true density threshold $\rho_0\!=\!48$ and 
    (c) magnitude threshold $th_{\rho_0}$.}
    \label{fig:MCCHMT2K25}
\end{figure}

We here discuss the dependence of the measure accuracy on the
coupling strength $C$ and use as example the system S1 with $K\!=\!25$ and
$n\!=\!2048$. In Fig.~\ref{fig:MCCHMT2K25}, the MCC for
PMIME, RGPDC, RCGCI and TE is given as a function of $C$ for the 
three significance criteria.
The parameters in the criteria are $\alpha\!=\!0.05$ for the
statistical testing, the true density of the original network $\rho_0\!=\!48$, 
and the average magnitude threshold $th_{\rho_0}$ over all magnitude thresholds
corresponding to the true density for the 10 realizations. For all significance
criteria, the PMIME exhibits the best performance for large
$C>0.1$ while TE has the highest MCC for small $C$ at 0.05 and
0.1. Though S1 is a nonlinear system, the linear measures RCGCI
and RGPDC (using the same dimension reduction step) are
competitive and as good as or better than TE for large $C$. 
Thus for this system and setting of $n$, $K$ and large $C$, 
the rate of indirect (false) couplings found by the nonlinear bivariate measure TE is as large as or larger
than the undetected nonlinear couplings from the linear measures RCGCI
and RGPDC. This indicates that linear measures with dimension reduction
may even perform better than nonlinear ones in settings of time series from nonlinear systems. 
It is noted that the coupling strength $C\!=\!0.05$ is very weak and the dimension
reduction methods PMIME, RCGCI and RGPDC find no significant
causal effects giving zero, which cannot change with any
significance criterion. On the other hand, the small TE values for
$C\!=\!0.05$ are still found significant at a good proportion giving
rather large MCC at the level of 0.8.
 
\subsection{Ranking of causality measures for each synthetic system}

We derive summary results of all measures at each system over all
coupling strengths $C$ and for different number of variables $K$
and time series lengths $n$ where applicable. For this, we use the
average score index $s_{i,j}$ for each measure $i$ at each system
$j$ as defined in Sec.~\ref{subsec:score}. In all results in this
section, the statistical significance testing for $\alpha\!=\!0.05$
has been used.

In Table \ref{tab:HMSCORE}, the average score $s_{i,j}$ for system
S1 (coupled H\'{e}non maps) is presented for the eight measures
scoring highest at each scenario combining $K\!=\!5$ and $K\!=\!25$ with
$n\!=\!512$ and $n\!=\!2048$.
\begin{table}[H]
\caption{The ranking of the eight best measures according to the
score index for system S1 of coupled H\'{e}non maps for all
scenarios of number of variables $K$ and time series length $n$.}
\centering
\begin{tabular}{p{25.5mm}p{5.5mm}p{25.5mm}p{5.5mm}p{25.5mm}p{5.5mm}p{25.5mm}p{5.5mm}}
\toprule
\multicolumn{4}{c}{$K\!=\!5$}                                                                                & \multicolumn{4}{c}{$K\!=\!25$}                                                                               \\
\midrule
\multicolumn{2}{c}{$n\!=\!512$}                          & \multicolumn{2}{c}{$n\!=\!2048$}                        & \multicolumn{2}{c}{$n\!=\!512$}                         & \multicolumn{2}{c}{$n\!=\!2048$}                         \\
\midrule
Measure                                     & Score& Measure                                    & Score& Measure                                    & Score& Measure                                    & Score \\
\midrule
\textbf{PMIME}($L\!=\!5$)           & 0.87 & \textbf{PMIME}($L\!=\!5$)          & 0.84 & \textbf{PMIME}($L\!=\!5$)          & 0.92 & \textbf{PMIME}($L\!=\!5$)          & 0.91  \\
TERV($m\!=\!3$)                             & 0.78 & \textbf{RGPDC}($p\!=\!5$,$\delta$)  & 0.77 & TE($m\!=\!2$)                              & 0.87 & \textbf{RGPDC}($p\!=\!5$,$\delta$)  & 0.83  \\
STE($m\!=\!3$)                              & 0.77 & \textbf{RGPDC}($p\!=\!5$,$\theta$)  & 0.76 & TE($m\!=\!3$)                              & 0.86 & \textbf{RGPDC}($p\!=\!5$,$\theta$)  & 0.83  \\
TE($m\!=\!3$)                               & 0.72 & \textbf{RCGCI}($p\!=\!5$)          & 0.76 & \textbf{RGPDC}($p\!=\!5$,$\beta$)   & 0.81 & \textbf{RCGCI}($p\!=\!5$)          & 0.82  \\
\textbf{RCGCI}($p\!=\!5$)           & 0.71 & \textbf{RGPDC}($p\!=\!5$,$\gamma$)  & 0.75 & \textbf{RGPDC}($p\!=\!5$,$\alpha$)  & 0.81 & \textbf{RGPDC}($p\!=\!5$,$\beta$)   & 0.82  \\
\textbf{RGPDC}($p\!=\!5$,$\alpha$)   & 0.71 & \textbf{RGPDC}($p\!=\!5$,$\beta$)   & 0.75 & \textbf{RGPDC}($p\!=\!5$,$\theta$)  & 0.79 & \textbf{RGPDC}($p\!=\!5$,$\alpha$)  & 0.81  \\
CGCI($p\!=\!5$)                             & 0.70 & PDC($p\!=\!5$,$\alpha$)                    & 0.75 & \textbf{RGPDC}($p\!=\!5$,$\delta$)  & 0.78 & \textbf{RGPDC}($p\!=\!5$,$\gamma$)  & 0.80  \\
\textbf{RGPDC}($p\!=\!5$,$\gamma$)   & 0.70 & GPDC($p\!=\!5$,$\alpha$)                   & 0.73 & \textbf{RCGCI}($p\!=\!5$)          & 0.77 & CGCI($p\!=\!5$)                            & 0.76  \\
\bottomrule
\end{tabular}
\label{tab:HMSCORE}
\end{table}
In all scenarios of S1 the PMIME measure is found to have the best
performance. Also, the other measures of dimension reduction RCGCI
and RGPDC reach highly ranked positions in all scenarios,
especially in the case of large time series length. It is noted
that these two measures are linear and they beat many other
nonlinear measures showing the importance of proper dimension
reduction. For small time series length ($n\!=\!512$), the information
measures show better performance and it is again notable that the
bivariate measures, such as TE, STE and TERV, score higher than
the corresponding multivariate measures, PTE PSTE and PTERV.
Again, the explanation for this lies in the inability of the
multivariate measures to deal with high dimensions if
dimension reduction is not employed. Having even as
low as three conditioning variables in the conditional mutual
information used by these measures (in the case of $K\!=\!25$ the
three more correlated variables in terms of MI to the driving
variable are selected from the 23 remaining variables) does not
provide as accurate estimates of the causal effects as the
respective bivariate measures. These multivariate measures (along
other multivariate measures of no dimension reduction) give
non-existing causal effects even to the beginning and end of the
ring, whereas the respective bivariate measures do not, and
only estimate additionally indirect causal effects (results not shown here).

In Table \ref{tab:MGSCORE}, the average score for system S2 of
coupled Mackey-Glass subsystems is presented for the eight
measures scoring highest at each scenario combining $K\!=\!5$ and
$K\!=\!25$ with $\Delta\!=\!100$ and $\Delta\!=\!300$, where $\Delta$ controls
the complexity of each subsystem.
\begin{table}[H]
\caption{The ranking of the eight best measures according to the
score index for system S2 of coupled Mackey-Glass subsystems for all
scenarios of number of variables $K$ and delay parameter $\Delta$ that controls the complexity of each subsystem.}
\centering
\begin{tabular}{p{25.5mm}p{5.5mm}p{25.5mm}p{5.5mm}p{25.5mm}p{5.5mm}p{25.5mm}p{5.5mm}}
\toprule
\multicolumn{4}{c}{$K\!=\!5$}                                                                                     & \multicolumn{4}{c}{$K\!=\!25$}                                                                                  \\
\midrule
\multicolumn{2}{c}{$\Delta\!=\!100$}                    & \multicolumn{2}{c}{$\Delta\!=\!300$}                      & \multicolumn{2}{c}{$\Delta\!=\!100$}                      & \multicolumn{2}{c}{$\Delta\!=\!300$}                  \\
\midrule
Measure                                     & Score & Measure                                       & Score & Measure                                       & Score & Measure                                   & Score \\
\midrule
\textbf{RGPDC}($p\!=\!20$,$\alpha$)  & 0.90  & \textbf{RCGCI}($p\!=\!20$)            & 0.88  & \textbf{PMIME}($L\!=\!50$)            & 1.00  & \textbf{PMIME}($L\!=\!50$)        & 0.95  \\
\textbf{RGPDC}($p\!=\!20$,$\delta$)  & 0.88  & \textbf{RGPDC}($p\!=\!20$,$\gamma$)            & 0.86  & PGCI($p\!=\!5$)                                & 0.87  & \textbf{RCGCI}($p\!=\!5$)         & 0.89  \\
\textbf{RGPDC}($p\!=\!20$,$\beta$)   & 0.88  & \textbf{RGPDC}($p\!=\!20$,$\delta$)    & 0.86  & \textbf{RGPDC}($p\!=\!20$,$\gamma$)    & 0.85  & \textbf{RGPDC}($p\!=\!5$,$\delta$) & 0.88  \\
\textbf{RGPDC}($p\!=\!20$,$\gamma$)  & 0.88  & \textbf{RGPDC}($p\!=\!20$,$\alpha$)    & 0.86  & \textbf{RGPDC}($p\!=\!20$,$\delta$)    & 0.85  & \textbf{RGPDC}($p\!=\!5$,$\gamma$) & 0.87  \\
\textbf{RCGCI}($p\!=\!20$)          & 0.86  & \textbf{RGPDC}($p\!=\!20$,$\theta$)    & 0.85  & \textbf{RGPDC}($p\!=\!20$,$\beta$)     & 0.84  & \textbf{RGPDC}($p\!=\!5$,$\beta$)  & 0.86  \\
\textbf{RGPDC}($p\!=\!20$,$\theta$)  & 0.86  & \textbf{RGPDC}($p\!=\!20$,$\beta$)    & 0.85  & \textbf{RCGCI}($p\!=\!20$)            & 0.84  & \textbf{RGPDC}($p\!=\!5$,$\alpha$) & 0.86  \\
\textbf{RGPDC}($p\!=\!5$,$\theta$)   & 0.86  & \textbf{RGPDC}($p\!=\!5$,$\gamma$)     & 0.80  & \textbf{RGPDC}($p\!=\!20$,$\alpha$)    & 0.83  & \textbf{RGPDC}($p\!=\!5$,$\theta$) & 0.85  \\
\textbf{RGPDC}($p\!=\!5$,$\delta$)   & 0.84  & \textbf{RCGCI}($p\!=\!5$)     & 0.78  & PGCI($p\!=\!20$)                               & 0.82  & \textbf{RGPDC}($p\!=\!20$,$\alpha$)& 0.81  \\
\bottomrule
\end{tabular}
\label{tab:MGSCORE}
\end{table}
This system is comprised of highly complex systems with complexity
increasing with $\Delta$. For $K\!=\!5$ and regardless of $\Delta$,
the linear measures using dimension reduction RCGCI and RGPDC show
the best performance, indicating again the importance of dimension
reduction, here for oscillating complex systems. The PMIME scores
slightly lower than these measures and given that RGPDC scores
equally high at different bands, together with RCGCI they occupy
the first eight places, so that the PMIME is not listed. For
$K\!=\!25$ on the other hand, the PMIME scores much higher than the
RCGCI and RGPDC measures and is at the first place for both
$\Delta\!=\!100$ and $\Delta\!=\!300$. Apparently, the dimension reduction
in the information measure of PMIME is more effective than in the
VAR-based measure of RCGCI and RGPDC for larger $K$.

In Table \ref{tab:NMSCORE}, the average score for system S3 of the
neural mass model is presented as for S2, but having as system
parameter $A\!=\!3.45, 3.7$, where the latter value indicates more
clear oscillating behavior.
\begin{table}[H]
\caption{The ranking of the eight best measures according to the
score index for system S3 of the neural mass model for all
scenarios of number of variables $K$ and oscillation controlling
parameter $A$.}
\centering
\centering
\begin{tabular}{p{25.5mm}p{5.5mm}p{25.5mm}p{5.5mm}p{25.5mm}p{5.5mm}p{25.5mm}p{5.5mm}}
\toprule
\multicolumn{4}{c}{$K\!=\!5$}                                          & \multicolumn{4}{c}{$K\!=\!25$}                                              \\
\midrule

\multicolumn{2}{c}{$A\!=\!3.45$}    & \multicolumn{2}{c}{$A\!=\!3.7$}        & \multicolumn{2}{c}{$A\!=\!3.45$}      & \multicolumn{2}{c}{$A\!=\!3.7$}           \\
\midrule

Measure            & Score    & Measure               & Score    & Measure              & Score    & Measure               & Score       \\
\midrule
\textbf{RCGCI}($p\!=\!20$)    & 0.88 & GPDC($p\!=\!20$,$\theta$)    & 0.94 & GPDC($p\!=\!5$,$\theta$)      & 0.83 & GPDC($p\!=\!20$,$\theta$)     & 0.92 \\
GPDC($p\!=\!20$,$\theta$)  & 0.88 & PDC($p\!=\!20$,$\theta$)     & 0.88 & GPDC($p\!=\!20$,$\theta$)   & 0.82 & CGCI($p\!=\!20$)       & 0.90 \\
\textbf{RGPDC}($p\!=\!20$,$\alpha$) & 0.88  & \textbf{PMIME($L\!=\!20$)} & 0.80 & \textbf{RCGCI}($p\!=\!20$)  & 0.81 & dDTF($p\!=\!20$,$\delta$)     & 0.89 \\
\textbf{RGPDC}($p\!=\!20$,$\gamma$) & 0.87 & \textbf{RGPDC}($p\!=\!5$,$\alpha$)    & 0.78 & \textbf{RGPDC}($p\!=\!20$,$\delta$)      & 0.79 & GPDC($p\!=\!5$,$\alpha$)       & 0.87 \\
PGCI($p\!=\!20$)    & 0.87 & \textbf{RCGCI}($p\!=\!5$)      & 0.78  & \textbf{RGPDC}($p\!=\!20$,$\beta$)    & 0.79 & \textbf{PMIME($L\!=\!20$)} & 0.87 \\
CGCI($p\!=\!20$)    & 0.87 & \textbf{RGPDC}($p\!=\!5$,$\alpha$)  & 0.77 & \textbf{RGPDC}($p\!=\!20$,$\theta$) & 0.79 & PDC($p\!=\!5$,$\alpha$)        & 0.82 \\
\textbf{RGPDC}($p\!=\!20$,$\theta$) & 0.86 & \textbf{RGPDC}($p\!=\!20$,$\theta$)      & 0.77 &  CGCI($p\!=\!20$)  & 0.78 & PDC($p\!=\!20$,$\theta$)      & 0.80 \\
\textbf{RGPDC}($p\!=\!20$,$\delta$) & 0.85 & \textbf{RGPDC}($p\!=\!20$,$\gamma$)    & 0.77 &  PDC($p\!=\!20$,$\theta$)      & 0.78  & GPDC($p\!=\!20$,$\alpha$)     & 0.80 \\
\bottomrule
\end{tabular}
\label{tab:NMSCORE}
\end{table}
In all scenarios GPDC shows the best performance. The RCGCI
measure for $A\!=\!3.45$ reaches the next position and also RGPDC
reaches a high position on the ranking in the first three
scenarios. The fact that GPDC scores higher than RGPDC also for
$K\!=\!25$ indicates that for this system and both $A$ the inclusion of all
lagged terms in VAR of order $p\!=\!5$ or $p\!=\!20$ gives somehow better 
identification of the correct couplings after significance testing. 
This is so, due to the relative large length $n\!=\!4096$ of the time series that
allows for the reliable estimation of the coefficients being as many as 
$20 \!\cdot\! 25 \!=\! 500$. 
For $A\!=\!3.45$, the PMIME does
not score high as its sensitivity is comparatively small (fails to
find significant proportion of true causal effects), whereas for
$A\!=\!3.7$ the PMIME is also among the first eight best measures. It
is observed that in all settings the frequency measures, and
particularly at low frequency bands, have the ability to identify
the true causality interactions better than information and other
measures. This is reasonable since this system is characterized by 
strongly harmonic oscillations.

For S4, no average results are shown as the system is run for only
one scenario, and the ranking for this was shown in
Table~\ref{tab:VARMCC} and discussed earlier.

\subsection{Overall ranking of causality measures}

For an overall evaluation of the causality measures, the average
score $s_i$ over all systems and scenarios is computed for each
measure $i$, as defined in Sec.~\ref{subsec:score}. In Table
\ref{tab:TOTALSCORE}, the ten measures with highest score $s_i$
are listed.
\begin{table}[H]
\caption{Average score index over all systems and scenarios.}
\centering
\begin{tabular}{cc}
\toprule
Measure & Score \\
\midrule
\textbf{PMIME}   & 0.80  \\
\textbf{RGPDC}    & 0.79  \\
\textbf{RCGCI}   & 0.78  \\
GPDC    & 0.63  \\
CGCI   & 0.61  \\
PGCI   & 0.61  \\
PDC    & 0.56  \\
TE     & 0.51  \\
dDTF    & 0.50  \\
TERV   & 0.46 \\
\bottomrule
\end{tabular}
\label{tab:TOTALSCORE}
\end{table}
It is noted that for each measure computed for varying parameters,
such as the frequency bands for the frequency measures, only the
one with the highest score is listed. The best performance is
achieved by the three measures making use of dimension reduction,
with the information measure PMIME scoring highest. It is noted
that there is a jump in score from the third to the fourth place,
showing the superiority of the measures of dimension reduction
over the other measures. The remaining places in the list are
dominated by the linear measures in the time and frequency domain.
Comparing the frequency measures based all on the same VAR model
we note that GPDC, and even PDC score higher than dDTF. As for the
information measures, the bivariate measures TE and TERV score
much higher than the multivariate respective measures (results not
shown), indicating the inability of multivariate information measures to
perform well unless an appropriate dimension reduction is applied.

\section{Discussion}
\label{sec:Discussion}

In this paper, a simulation study is performed for the estimation
of causality networks from multivariate time series. For the
network construction, Granger causality measures,
simply termed here as causality measures, of different type were
employed as information and model-based measures, measures based on phase,
frequency measures and measures making use of dimension reduction.
These measures are applied to linear and nonlinear (chaotic), deterministic
and stochastic, coupled simulated systems, to evaluate their ability to detect the
existing coupled pairs of observed variables of these systems. We
considered the nonlinear dynamical systems of coupled H\'{e}non
maps (S1), coupled Mackey-Glass subsystems (S2), the so-called
neural mass model (S3), and a linear vector autoregressive process
(VAR) of order 3 (S4). For systems S1, S2, S3 we used $K\!=\!5$ and
$K\!=\!25$ subsystems, whereas S4 was defined only on $K\!=\!25$
variables. For S2 and S3, we considered two regimes of different
complexity for each system, controlled by a system parameter. For
S1, S2, and S3, a range of coupling strengths $C$ were designed
covering the setting of none to weak and strong coupling. For S1,
a small and a large time series length $n$ were used. This design
of the simulation aimed at testing the causality measures on
different types of systems with respect to time (S1, S4 are
discrete and S2, S3 continuous in time), low and high dimensional
having $K\!=\!5$ and $K\!=\!25$, linear (S4) and nonlinear (S1, S2, S3),
deterministic (S1, S2) and stochastic (S3, S4), and for a range of
coupling strengths (S1, S2, S3). Thus, rather than concentrating
on a particular system or type of systems, e.g. often met in EEG
studies, we aimed at evaluating the performance of the causality
measures on many different system settings. Measures that were
best suited for strongly oscillating systems, such as the
frequency measures, may not be appropriate for maps, and on the
other hand, information measures on ranks (such as TERV) that are
more appropriate for discrete-time systems (maps) may not be
appropriate for strongly oscillating signals. However, the
evaluation showed that this was not the case, and, in the overall
ranking, frequency measures dominated, but also TERV was included
among the ten best.

The evaluation of the measures was based on the match of the
causality network constructed from each measure to the original
coupling network of the system generating the multivariate time
series. For this, three significance criteria were used to
transform the value of each causality measure, corresponding to a
weighted network connection, to a binary value, corresponding to a
binary connection. While the criteria of network density threshold
and magnitude threshold are arbitrary and best results are only
attained when the thresholds are given based on the knowledge of
the original coupling network, the statistical testing, which does
not require a-priori information on the underlying system, was
competitive and was further suggested as the criterion of choice
to derive binary connections. Performance indices were computed
checking the preservation of the original binary connections
(true) in the causality binary network (estimated), and we used
the Matthews correlation coefficient (MCC) to quantify best the
matching performance of each causality measure as it weighs
sensitivity and specificity.

We considered bivariate and multivariate causality measures, and a
subset of three multivariate measures making use of dimension
reduction. The first is the information measure of partial mutual
information from mixed embedding (PMIME), which can be considered
as a restriction of the partial transfer entropy (PTE) to the most
relevant lagged variables. The other two measures are linear and
they are both based on VAR model. The dimension reduction suggests
fitting a sparse VAR rather than a full VAR. While the conditional
Granger causality index (CGCI) is defined in the time domain on
the full VAR, the restricted CGCI (RCGCI) is computed on the
sparse VAR, and accordingly in the frequency domain the
generalized partial directed coherence (GPDC) is modified using a
sparse VAR to the restricted GPDC (RGPDC). 

While linear models can 
be estimated sufficiently well in high-dimensional time series, provided
the length of the time series is much larger than the number of the unknown model 
coefficients, the estimation of entropies, used in information measures,
in high dimension is problematic, even when using the most appropriate 
estimate of nearest neighbors. To make a fair
comparison between the multivariate information measure making use of
dimension reduction (PMIME) to the other multivariate information measures in
high-dimensional time series (here $K\!=\!25$), for the latter
measures we do not condition the causal relationship among the
driving and response to all remaining variables (23 in our case), but rather
select the three variables that are best correlated in terms of mutual
information to the driving variable. In this way we avoid to some
degree the curse of dimensionality, but still the embedding is
done separately for each of the five variables, i.e. the driving,
the response and the other three variables, whereas for the
measures using dimension reduction, the embedding is built jointly
for all variables selecting only the most appropriate lagged
variables.

The evaluation of the causality measures showed differences in
their performance in the different systems and their parameters
($n$, $K$, $C$ and system parameter $\Delta$ for S2 and $A$ for
S3). For system S1, the measures making use of dimension reduction
scored highest regardless of $n$ and $K$ with the PMIME attaining
highest MCC for all but very small $C$. Bivariate information
measures scored high here but only for small $n$. For system S2,
the frequency measures were the most accurate at all frequency
bands, especially for small $C$ while for stronger couplings
dimension reduction measures reached higher MCC. For
high-dimensional time series ($K\!=\!25$) again the PMIME scored
highest followed mainly by the linear dimension reduction measures
for different parameter values. For system S3 characterized by
strong oscillations, the frequency measures performed best
occupying the highest ranks and only the PMIME entered the list of
eight highest rankings for the system parameter $A\!=\!3.7$ for both
$K$. For the linear VAR system S4, as expected, the linear measures
with dimension reduction performed best and the respective linear
measures of full dimension had also increased performance.

The conclusions of the simulation study on comparatively
low-dimensional ($K\!=\!5$) and high-dimensional ($K\!=\!25$) time series
from different systems are itemized as follows:
\begin{enumerate}[leftmargin=*,labelsep=4.9mm]
\item The multivariate measures making use of dimension
    reduction (PMIME, RCGCI, RGPDC) outperform all other bivariate and multivariate
    measures.
    \item Among the dimension reduction measures, the information
    measure of PMIME is overall best but the overall score is
    slightly higher than that of the other two linear measures.
    Though the PMIME outperforms the other measures in the
    chaotic systems S1 and S2, for the strongly oscillating stochastic system
    S3 and the linear stochastic process S4 it scores lower than
    RCGCI and RGPDC.
    \item Linear measures, especially these applying dimension
    reduction, exhibited a competitive performance to other
    nonlinear measures also on nonlinear systems, such as S1, S2
    and S3. This remark supports the use of linear measures
    (preferably with dimension reduction) to settings that may
    involve nonlinear relationships. Certainly, results still
    depend on the studied system.
    \item Though bivariate measures tend to identify causality
    relationships that are not direct, they do not fail in
    identifying non-existing causal relationships. The latter occurs
    when using multivariate measures without dimension reduction. Though this
    effect cannot be captured by standard performance indices used in this study,
    such as the sensitivity and specificity, it is a
    significant finding advocating against the use of multivariate
    measures, unless dimension reduction is applied.
\end{enumerate}

Admittedly, the collection of causality measures is biased
including all measures our team has developed. On the other hand,
the collection is not comprehensive, leaving out nonlinear
measures that are more difficult to implement and could not be
found freely available when the study was initiated. It is noted
that initially, many connectivity measures that are not
directional, especially these based on phases, were included, but
they could not be fairly evaluated in the designed framework
comparing the derived network to the original network of directed
connections. The simulation study was conducted using four
systems, leaving out other systems, such as the coupled Lorenz or
coupled R{\"o}ssler systems, as well as different coupling
structures, such as the random (used here only in S4), small-world
and scale-free, used by our team in other studies. Besides these
shortcomings, we believe the current study can be useful for
methodologists and practitioners to assess the strengths and
weaknesses of the different causality measures and their
applicability especially to high-dimensional time series.

\funding{Part of the research was supported by a grant from the Greek General Secretariat for Research and Technology
(Aristeia II, No 4822).}




\reftitle{References} 
\externalbibliography{yes}



\end{document}